\documentclass{aa}  

\newcommand{\AAA}{\boldsymbol{A}}
\newcommand{\BB}{\boldsymbol{B}}
\newcommand{\JJ}{\boldsymbol{J}}
\newcommand{\ee}{\boldsymbol{e}}
\newcommand{\kk}{\boldsymbol{k}}
\newcommand{\xx}{\boldsymbol{x}}
\newcommand{\bnabla}{\boldsymbol{\nabla}}
\newcommand{\ii}{\mathrm{i}}
\newcommand{\bra}[1]{\langle #1\rangle}

\newcommand{\EEq}[1]{Equation~(\ref{#1})}
\newcommand{\Eq}[1]{Eq.~(\ref{#1})}
\newcommand{\Tab}[1]{Table~\ref{#1}}

\newcommand{\Sec}[1]{Sect.~\ref{#1}}
\newcommand{\Eqs}[2]{Eqs.~(\ref{#1}) and~(\ref{#2})}

\newcommand{\xics}{\xi_\text{cs}}
\newcommand{\etacs}{\eta_\text{cs}}
\def\EM{E_{\rm M}}
\def\EA{E_{\rm A}}
\def\HM{H_{\rm M}}
\def\Lu{\mbox{\rm Lu}}
\newcommand{\G}{\,{\rm G}}
\newcommand{\Myr}{\,{\rm Myr}}
\newcommand{\km}{\,{\rm km}}

\usepackage{color}
\definecolor{blue}{rgb}{0.15, 0.38, 0.61}

\usepackage{amsmath}
\usepackage{graphicx}
\usepackage{txfonts}
\usepackage{hyperref}

\begin{document}

\title{Reality of inverse cascading in neutron star crusts}

\author{Clara Dehman
\inst{1}\fnmsep\inst{2}\thanks{Email: clara.dehman@ua.es} \and Axel Brandenburg\inst{2,3,4,5}  }

\institute{Departament de F\'isica Aplicada, Universitat d'Alacant, Ap.\ Correus 99, E-03080 Alacant, Spain 
\and
Nordita, KTH Royal Institute of Technology and Stockholm University, 10691 Stockholm, Sweden
\and 
The Oskar Klein Centre, Department of Astronomy, Stockholm University, AlbaNova, SE-10691 Stockholm, Sweden
\and
McWilliams Center for Cosmology \& Department of Physics, Carnegie Mellon University, Pittsburgh, PA 15213, USA
\and 
School of Natural Sciences and Medicine, Ilia State University, 3-5 Cholokashvili Avenue, 0194 Tbilisi, Georgia
}

\date{}

\abstract{
The braking torque that dictates the timing properties of magnetars is closely tied to the large-scale dipolar magnetic field on their surface.
The formation of this field has been a topic of ongoing debate.
One proposed mechanism, based on macroscopic principles, involves an inverse cascade within the neutron star's crust.
However, this phenomenon has not been observed in realistic simulations. In this study, we provide compelling evidence supporting the feasibility of the inverse cascading process in the presence of an initial helical magnetic field within realistic neutron star crusts and discuss its contribution to the amplification of the large-scale magnetic field. 
Our findings, derived from a systematic investigation that considers various coordinate systems, peak wavenumber positions, crustal thicknesses, magnetic boundary conditions, and magnetic Lundquist numbers, reveal that the specific geometry of the crustal domain—with its extreme aspect ratio—requires an initial peak wavenumber from small-scale structures for the inverse cascade to occur.
However, this same aspect ratio confines the cascade to structures on the scale of the crust, making the formation of a large-scale dipolar surface field unlikely.
Despite these limitations, the inverse cascade remains a significant factor in the magnetic field evolution within the crust and may help explain highly magnetized objects with weak surface dipolar fields, such as low-field magnetars and central compact objects.
}


\maketitle

\section{Introduction} 
\label{sec: intro}
Magnetars are the most magnetic objects within the population of neutron stars (NSs) \citep{turolla2015, esposito2021}. Their bright X-ray luminosity and occurrence of observed bursts and outbursts are attributed to the restless dynamics and dissipation of a strong magnetic field (about $10^{14}$ to $10^{15}$\,G) near the NS surface \citep{thompson1995}. Moreover, the large-scale dipolar field on the surface regulates the braking torque responsible for their timing properties \citep{ostriker1969}, resulting in long spin-periods.

The origin and evolution of NS magnetic fields have been long-debated topics. While it is widely accepted that the fossil magnetic field inherited from the progenitor star is insufficient to account for the strongest observed NS fields, additional amplification is necessary. This amplification might occur during the proto-NS stage through a turbulent dynamo process \citep{balbus1991,obergaulinger2014,raynaud2020,reboul2021,aloy2021}. 
Magnetic field amplification can also take place later in an NS's life, 
for instance due to the re-emergence of a buried magnetic field driven by Hall drift (see \citet{igoshev2021} for a review), or through the chiral magnetic instability (CMI) in the NS crust, a microscopic-based mechanism that can explain the formation of large-scale magnetic fields \citep{dehman2024}.

Understanding the amplification of the surface dipolar field in magnetars is a challenging task, where macroscopic mechanisms like inverse cascading could be important phenomena to consider.
A recent attempt to explain this phenomenon, utilizing an initial turbulent field structure derived from proto-NS dynamo simulations \citep{reboul2021}, is presented in \citet{dehman2023b}. The authors conducted the first 3D coupled magneto-thermal simulation in the crust of a NS, incorporating temperature-dependent microphysical calculations and a realistic stellar structure. The study aimed to determine if starting from such an initial field configuration could explain the observed dipolar field in magnetars. While successful in explaining the characteristic energy spectrum of low-field magnetars, high-field pulsars, and central compact objects (CCOs), this study falls short in explaining the strong surface dipolar field observed in magnetars.
The surface dipolar component remained constant over time, with no evidence of an inverse cascade.
Other simulations dedicated to the investigation of CCOs have also sought to observe inverse cascading, as seen in \citet{gourgouliatos2020}.
Although they did find a substantial increase of the dipole component of the magnetic field,
the location of the peak of the spectrum remained unchanged.
Because of this, they concluded that their simulations showed no signs of a true inverse cascade.

Earlier work by \citet{WH09,WH10} and \citet{Cho11} identified the role of magnetic helicity in driving an inverse cascade under the influence of Hall drift.
An inverse cascade was known to exist in magnetohydrodynamics with helical driving \citep{frisch1975}.
Building on those findings,
recent local simulations of the Hall cascade with initial magnetic helicity were presented by \citet{brandenburg2020}.
This model was later applied to simulate the evolution of a population of NSs \citep{sarin2023}.
In the study by \citet{brandenburg2020}, the author proposed a model suggesting that the large-scale magnetic field of NSs grows as a result of small-scale turbulence.
Initially, the magnetic field in young NSs may predominantly exist at small scales, but it could later undergo an inverse cascade, particularly after the crust solidifies. 
This process implies that the spectral magnetic energy at lower multipoles would increase over time rather than decrease.
The study further demonstrates that the resulting dipolar field intensifies approximately linearly with time, typically growing by three orders of magnitude, while thermal dissipation gradually diminishes.

As explained by \citet{brandenburg2020}, the presence of an initial helical magnetic field could be a key factor behind the observed phenomenon of inverse cascading. The simulations further suggest that, even with a relatively weak initial magnetic field, the occurrence of strong inverse cascading is evident. There are two possible sources of magnetic helicity in a NS.
One is through dynamo action by neutrino-driven convection in young NS, resulting in oppositely signed magnetic helicities in the two hemispheres \citep{thompson1995,brandenburg2005}. 
Another possible origin of helical magnetic fields in NSs is the CMI. This effect is significant during supernova explosions, throughout the proto-NS phase, and in young NSs. The CMI influences the evolution of the magnetic field in these compact objects, leading to the saturation of magnetic helicity. This, in turn, results in the growth of magnetic fields, which is governed by a conservation law for total chirality \citep{sigl2016,rogachevskii2017, schober2018, masada2018, brandenburg2023a,
brandenburg2023b,dehman2024}. Consequently, adopting an initial helical magnetic field provides a realistic foundation for modeling the long-term evolution of NS magnetic fields.

In addition to the significance of the initial helical field in explaining the phenomenon of inverse cascading, several other factors play a crucial role in comprehending the occurrence of inverse cascading within a realistic NS crust. Building upon the findings of \citet{brandenburg2020}, where inverse cascading was first observed in the NS crust, several extensions can be explored. These include incorporating a spherical shell geometry, accounting for the precise aspect ratio determined by applying the Tolman-Oppenheimer-Volkoff equation \citep{oppenheimer1939} and employing a realistic nuclear matter equation of state (EoS). Furthermore, instead of imposing periodic boundary conditions, alternative boundary conditions can be considered. Options such as potential boundary conditions at the surface and perfect conductor at the crust-core interface may provide valuable insights. However, at the surface of the star, the most suitable boundary conditions are force-free ones \citep{akgun2018,urban2023}. Finally, within the NS, the stratification of matter is a crucial consideration. The inclusion of density- and temperature-dependent microphysics becomes imperative for accurately computing the Hall prefactor and magnetic diffusivity employed in the evolution equations.

In this article, we aim to explore the potential occurrence of inverse cascading in the crust of a NS by simulating a realistic NS scenario. 
Our objective is to understand the role of inverse cascading in explaining the strong surface dipole field observed in magnetars.
Additionally, we seek to address why this phenomenon has not been identified in previous studies within the NS community.
We investigate this process over a time span of $10^3$ to $10^5$ years, corresponding to the observed lifetime of magnetars.
To achieve these goals, we employ both the \textsc{Pencil Code}\footnote{\url{https://github.com/pencil-code}} \citep{pencil2021} and \texttt{MATINS} \citep{dehman2023a}, taking into account the various relevant parameters. 

This paper is organized as follows: In \Sec{sec: helicity}, we discuss the theoretical formalism of magnetic helicity and its realizability conditions in both Cartesian and spherical coordinates. The numerical setups and the initial conditions
are described in \Sec{sec: numerical setup}. The results of our simulations are presented in \Sec{sec: Simulations}, and we conclude with a discussion of our findings in \Sec{sec: discussion}.

\section{Magnetic helicity and realizability condition}
\label{sec: helicity}

\subsection{General considerations}

As previously noted, a helical magnetic field plays an important role in explaining inverse cascading, which might be considered as a candidate to explain the large-scale dipolar field in magnetars. The idea of explaining inverse cascading using a helical field was first
proposed by \citet{frisch1975}. The first application to NS physics was explored by \citet{brandenburg2020}. The idea is based on the conservation of magnetic helicity,
\begin{equation}
    \HM = \int_V \AAA \cdot \BB \, dV, 
    \label{eq: magnetic helicity}
\end{equation}
where $\AAA$ is the magnetic vector potential and $\BB= \bnabla \times \AAA$ is the magnetic field in a volume $V$. 

Following \citet{brandenburg2020}, simple coordinate-independent measures
related to magnetic helicity can be defined in terms of the fractional
helicity $\chi$ and the length scale $\xi$, which are introduced through
the ratios
\begin{equation}
\bra{\AAA\cdot\BB}/\bra{\BB^2}\equiv\chi\xi,\quad
\bra{\AAA\cdot\BB}/\bra{\JJ\cdot\BB}\equiv\mu_0\xi^2.
\end{equation}
Here, $\JJ=\bnabla\times\BB/\mu_0$ is the current density, where $\mu_0$ is the vacuum permeability (in practice, we often use units such that $\mu_0=1$).
Angle brackets denote averages over closed volumes, so $\bra{\AAA\cdot\BB}$ is gauge-invariant.
These relations yield $\chi^2=\mu_0\bra{\AAA\cdot\BB}\bra{\JJ\cdot\BB}/\bra{\BB^2}^2$,
from which the fractional magnetic helicity $\chi$ is determined.

We recall that for a fully helical Beltrami field with periodic boundary conditions, we have $\chi=1$.
When the boundary conditions are nonperiodic, for example perfectly conducting on one side and a vertical field condition on the other,
\citet{brandenburg2017a} determined analytically the quantity $\epsilon_\mathrm{m}\equiv\bra{\JJ\cdot\BB}/(\bra{\JJ^2}\bra{\BB^2})^{1/2}\approx0.883$,
which also implies $\chi=0.883$, because in his case $\bra{\JJ\cdot\BB}/\bra{\AAA\cdot\BB}=\mu_0\bra{\JJ^2}/\bra{\BB^2}$.
Thus, $\chi$ can be close to unity even in the nonperiodic case.
Furthermore, because $\chi$ is only an approximate quantity, it can sometimes also exceed unity by a certain amount \citep{brandenburg2020}.

The $\chi$ approach provides a practical alternative to more rigorous methods that are typically feasible in Cartesian coordinates, which will be discussed next.
Using $\chi$, when treated as position-dependent, is also a useful tool if the magnetic diffusivity and/or the Hall prefactor vary with depth \cite[see][for examples]{brandenburg2020}.
However, in order to clarify the viability of an inverse cascade in spherical geometry, such variations will be ignored in the present paper.

\subsection{Cartesian coordinates}
\label{subsec: cart coord}
Assuming periodic boundary conditions, the spectra of magnetic energy and magnetic helicity, $\EM(k)$ and $\HM(k)$, can be computed by calculating the three-dimensional Fourier transforms of 
the magnetic vector potential $\boldsymbol{\hat{A}_k}$ and the magnetic field $\boldsymbol{\hat{B}_k}$.
The spectra are obtained by integrating $|\boldsymbol{\hat{B}_k}|^2$ and the real part of $\boldsymbol{\hat{A}_k}\cdot \boldsymbol{\hat{B}_k}^*$ over shells of constant $k=|\boldsymbol{k}|$, yielding $\EM(k)$ and $\HM(k)$, respectively.
Here, the asterisk denotes complex conjugation.
These spectra are normalized such that $\int \EM(k)\, dk = \bra{\BB^2}/2\mu_0$ and $\int \HM(k)\, dk = \bra{\AAA \cdot \BB}$ for $k$ from $0$ to $\infty$.
By applying the Schwartz inequality, one can derive the so-called realizability condition \citep{Mof78},
\begin{equation}
    k |H_M(k)|/2\mu_0\leq E_M(k).
    \label{eq: realisability cond}
    \end{equation}
The realizability condition serves as an indicator of the degree of helicity within the magnetic field. When \Eq{eq: realisability cond} is saturated 
at a particular wavenumber $k$, the magnetic helicity is maximal at that wavenumber. If this condition holds true for any arbitrary $k$, 
the system is characterized as being in a state of maximal helicity \citep{frisch1975}.

For fully helical magnetic fields with (say) positive helicity (e.g., $H_M = 2 \mu_0 E_M (k)/k$), \citet{frisch1975} demonstrated that energy and magnetic helicity cannot cascade directly. 
Specifically, the interaction of modes with wavenumbers $\boldsymbol{p}$ and $\boldsymbol{q}$ produces fields whose wavevector $\boldsymbol{k} =\boldsymbol{p}+\boldsymbol{q}$ has a length that is equal to or smaller than the maximum of either $|\boldsymbol{p}|$ or $|\boldsymbol{q}|$ \citep{brandenburg2005}, expressed as  
\begin{equation}
    |\boldsymbol{k}| \leq \mathrm{max}(|\boldsymbol{p}|,|\boldsymbol{q}|). 
    \label{eq: wavevector fully helical}
\end{equation} 
This implies that magnetic helicity and magnetic energy transform into progressively larger length scales, defining what is known as the inverse cascade. Consequently, the entire spectrum is expected to shift leftward, toward larger length scales, in an approximately self-similar fashion \citep{BK17}.  

For the decay of a fully helical magnetic field, the height of the spectrum $\EM^{\max}$ remains unchanged.
This is because the typical length scale $\xi$ increases
($k\propto1/\xi$ decreases for a fully helical field; see \Eq{eq: wavevector fully helical}),
but the mean magnetic helicity density $\bra{\AAA\cdot\BB}\approx \xi \bra{\BB^2}$ is constant,
so the mean magnetic energy density $\bra{\BB^2}/2\mu_0\approx\EM^{\max}/\xi$ must decrease such that $2\mu_0\EM^{\max}\approx\bra{\AAA\cdot\BB}$ stays constant.
Thus, the constancy of the height of the spectrum in the case of a fully helical field does not indicate that the mean magnetic energy density is constant during the evolution.  

\subsection{Spherical coordinates}
\label{subsec: spherical coordinates}
In spherical coordinates, $\BB$ can be expressed using two scalar functions $\Phi(\xx)$ and $\Psi(\xx)$, following the
Chandrasekhar-Kendall formulation \citep{chandrasekhar1957,chandrasekhar1981}:
  \begin{eqnarray}
      \BB_\text{pol} &=&  
    \bnabla \times \big( \bnabla \times \Phi \boldsymbol{r} \big),
      \nonumber\\
        \BB_\text{tor} &=& 
        \bnabla\times \Psi \boldsymbol{r}.
        \label{eq: poloidal toroidal field components}
  \end{eqnarray}
Using the notation of \citet{KR80} and \citet{geppert1991}, the basic idea is to expand the poloidal $\Phi$ and toroidal $\Psi$ scalar functions in a series of spherical harmonics:
  \begin{eqnarray}
    \Phi(t,r,\theta,\phi) &=& \frac{1}{r}\sum_{\ell,m} \Phi_{\ell m}(r,t) Y_{\ell m}(\theta,\phi),  \nonumber\\
    \Psi(t,r,\theta,\phi)  &=& \frac{1}{r}\sum_{\ell,m} \Psi_{\ell m}(r,t) Y_{\ell m}(\theta,\phi), 
       \label{eq: phi and Psi scalar functions}
  \end{eqnarray}
where $\ell= 1$ to $\infty$ is the degree and $m=-\ell,...,\ell$ the order of the multipole. The toroidal field in 3D is a mix of the two tangential components of the magnetic field, whereas the poloidal field is a mix of all three components.
This is less trivial than in 2D, where the toroidal part consists of the azimuthal component and the poloidal part consists of the two other components of the magnetic field.

The spectra of the magnetic helicity and the magnetic energy in 3D spherical coordinates are, respectively, defined as follows:
\begin{equation}
        \HM(\ell,m;t) = 2 \int \frac{\ell(\ell+1)}{r^2}  \Phi_{\ell m} \Psi_{\ell m}  r^2 dr,
        \label{eq: spectral magnetic helicity}
\end{equation}
\begin{equation}
        \EM(\ell,m;t) = \frac{1}{2} \int \frac{\ell(\ell+1)}{r^2} \bigg[  \frac{\ell(\ell+1)}{r^2} \Phi_{\ell m}^2 +\Phi^\prime_{\ell m} \,^2   +\Psi_{\ell m}^2  \bigg]. 
        \label{eq: energy spectrum MATINS}
\end{equation}
Here, $\Phi^\prime_{\ell m} = \partial \Phi_{\ell m}/\partial r$. From these expressions, one can define the spectral realizability condition (\Eq{eq: realisability cond}) in terms of the poloidal and toroidal scalar functions as
\begin{equation}
     2 k\, \Phi_{\ell m} \Psi_{\ell m} \leq  \frac{\ell(\ell+1)}{r^2} \Phi_{\ell m}^2 + \Phi^\prime_{\ell m}\,^2 + \Psi_{\ell m}^2.
    \label{eq: realisability l}
\end{equation}
The pseudo wavenumber $k=\sqrt{\ell(\ell+1)}/R$ possesses the dimension of inverse length, where $R$ denotes the surface of our computational domain.

\section{Numerical models}
\label{sec: numerical setup}
\subsection{Evolution equations} 
\label{sec:Lu}
Using SI units, the equations governing the Hall cascade in NS crust can be written as \citep{GR92}
\begin{equation}
    \frac{\partial \BB}{\partial t} =  \bnabla \times \bigg( - \frac{\JJ \times \BB}{e n_e} - \eta \mu_0 \JJ \bigg)\,, 
    \label{eq: induction equation}
\end{equation}
where $e$ is the unit charge, $n_e$ is the electron density, 
and $\eta$ is the magnetic diffusivity (inversely proportional to the electrical conductivity $\sigma_e$). 

A suitable adimensional measure quantifying the importance of magnetic diffusivity is the Lundquist number, given by
\begin{equation}
\Lu = B_\text{rms}/e n_e \mu_0\eta.
\label{eq: Lu}
\end{equation}
Owing to the decay of $B_\text{rms}$, the value of $\Lu$ is time-dependent.
Additionally, it exhibits positional dependence because $n_e$ and $\eta$ vary with density. However, this positional dependence is not addressed in the present study.

\subsection{Numerical codes}
\label{sec: numerics}
As mentioned above, we use two numerical codes: the
\textsc{Pencil Code}\footnote{\url{https://github.com/pencil-code}} \citep{pencil2021} and \texttt{MATINS} \citep{dehman2023a}. The \textsc{Pencil Code} is primarily designed to solve the fully nonlinear, compressible hydromagnetic equations.
Its highly modular structure allows for easy adaptation to a wide range of physical setups.
It is a high-order finite-difference code, which is efficiently parallelized,
enabling therefore high resolution and Lundquist numbers on the order of $10^3$.
The code can operate on a Cartesian or spherical grid and can be configured to work within a limited sector, $\theta_1\leq\theta\leq\theta_2$ and $\phi_1\leq\phi\leq\phi_2$, of a 3D spherical shell. 
This capability allows us to exclude the axis to avoid singularities and to adjust the aspect ratio as needed. Therefore, the \textsc{Pencil Code} is a valuable tool for this study. 

On the other hand, \texttt{MATINS} is a 3D code for the MAgneto-Thermal evolution in Isolated NS crusts, based on finite-volume numerical schemes discretized over a nonorthogonal cubed-sphere grid, which effectively resolves the axis singularity problem in 3D spherical coordinates. The cubed-sphere formalism, introduced by \citet{ronchi96} and implemented in \texttt{MATINS} by \citet{dehman2023a}, uses one coordinate as the radial direction, similar to spherical coordinates, with the volume composed of multiple radial layers. Each layer is divided into six patches, resembling arcs of great circles, created by inflating the six faces of a cube into a spherical shape. Each patch is bordered by four others and is described by angular-like coordinates, $\xics$ and $\etacs$, ranging from $[-\pi/4:\pi/4]$. These patches are orthogonal to the radial direction but not orthogonal to each other, except at the patch centers. This nonorthogonality requires careful distinction between covariant (lower indices) and contravariant (upper indices) field components.

Additionally, \texttt{MATINS} incorporates a spherical star based on a realistic EoS, including corresponding relativistic factors in the evolution equations. It also integrates the latest temperature-dependent microphysical calculations\footnote{The public routines implemented in \texttt{MATINS} are available at \url{http://www.ioffe.ru/astro/conduct/}.  For details, see \citet{potekhin2015}.}, enabling coupled magneto-thermal simulations if needed \citep{dehman2023a,dehman2023b,ascenzi2024}.

\subsection{Initial conditions}
\label{sec: initial conditions}
In the following, our objective is to construct an initial helical magnetic field. We use a random initial field with a specific magnetic energy spectrum. It peaks at a certain wavenumber $k_0$. For wavenumbers $k>k_0$, the spectrum exhibits a distinct inertial range,
which may follow a $k^{-5/3}$ scaling for Kolmogorov-like turbulence or a $k^{-2}$ scaling for wave turbulence \citep{BKT15}. For $k<k_0$, referred to as the sub-inertial range, we adopt a spectrum corresponding to a random vector potential. This implies that in 3D space, the vector potential follows a $k^2$ spectrum and the magnetic field a $k^4$ spectrum, as depicted in \Tab{tab:initial spectrum}. Such a spectrum is often used in the cosmological context, where it is usually referred to as a causal spectrum \citep{DC03}. It means that no point is correlated with any other, but the field is additionally divergence-free.

\begin{table}
    \caption{Spectrum of the magnetic vector potential $\AAA$ and the magnetic field $\BB$ for $k \ll k_0$, depicting the ascending spectra.  }
    \centering
    \begin{tabular}{ccc}
          & $\EA(\kk)$& $\EM(\kk)$\\
         \hline
      1D   & $\propto k^0$  & $\propto k^2$\\
       2D   & $\propto k^1$  & $\propto k^3$\\
        3D   & $\propto k^2$  & $\propto k^4$\\
    \end{tabular}
    \tablefoot{For more details, see \citet{BB2020}.}
    \label{tab:initial spectrum}
\end{table}

\subsubsection{The Pencil Code}
\label{sec: pencil code initial conditions}
In Cartesian coordinates, we use a Fourier transform to construct a helical initial condition for the magnetic vector potential $\AAA(\kk)$
by applying the helicity operator $R_{ij}(\kk)=\delta_{ij}-\ii\sigma\varepsilon_{ijl}\hat{k}_l$ with unit vector $\hat{\kk}=\kk/|\kk|$ on a nonhelical transverse field given by:
\begin{equation}
\AAA=A_0\frac{\kk\times\ee}{|\kk\times\ee|} S_A(k) \, e^{\ii\varphi}.
\end{equation}
Here, $A_0$ is an amplitude factor, $\varphi$ with $|\varphi|<\pi$
are uniformly distributed random phases, and
\begin{equation}
    S_A(k)= \frac{k_0^{-3/2} (k/k_0)^{\alpha/2-2} }{\big[ 1+ (k/k_0)^{2(\alpha+7/3)}  \big]^{1/4}}.
\end{equation}
is a function that gives $\EM(\kk)\propto k^\alpha$ for $k\ll k_0$ and $\EM(\kk)\propto k^{-7/3}$ for $k\gg k_0$. For more details on how $\alpha$ scales with the dimension of the
domain, we refer to \Tab{tab:initial spectrum} and to the work of \citet{BB2020}.
Here, $k_0$ is the peak wavenumber of the spectrum.
For a given value of $B_0$, the resulting initial value of the rms magnetic field, $B_\text{rms}$, which will be denoted as $B_\text{rms}^{(0)}$,
is usually somewhat larger than $B_0$.
For $k_0/k_1=180$, for example, we find $B_\text{rms}^{(0)}/B_0\approx 3.2$ when $\sigma=0$, and $B_\text{rms}^{(0)}/B_0\approx 4.5$ when $\sigma=1$.

Owing to the use of Fourier transforms, our initial conditions are implicitly assumed to be triply periodic in $\xx$, where $\xx=(x,y,z)$ with $x_1<x<x_2$, $y_1<y<y_2$, and $z_1<z<z_2$.
The boundary conditions imposed during the simulation may lead to sharp gradients at the boundaries during the initial time steps, but those will be smoothed out later in time.

When using the \textsc{Pencil Code}, we can still use the Cartesian initial condition in spherical geometry by replacing $(x,y,z)\to(r,\theta,\phi)$.
In that case, the vector potential will no longer be perfectly divergence-free.
Also, in the simulations, we are not working in the Coulomb gauge;
instead, we employ the Weyl gauge where the electrostatic potential vanishes.
Additionally, the fractional magnetic helicity will be different in this scenario (see \Sec{sec: helicity} for coordinate-independent measures of the fractional magnetic helicity).

In the following, we characterize the surface magnetic field in the Cartesian simulations through its two-dimensional energy spectrum, $\EM(k;x,t)$, where $k=|\kk|$ and $\kk=(k_y,k_z)$ is the wavevector in the $yz$ plane. We choose a value of $x$ that is close to the outer boundary, $x_2$ (beneath the star's surface). Owing to the reduction in dimensionality, the otherwise $k^4$ spectrum of the magnetic field turns into a $k^3$ spectrum (see \Tab{tab:initial spectrum}).

\subsubsection{The \texttt{MATINS} code}
\label{sec: MATINS initial conditions}
According to the Chandrasekhar-Kendall formalism \citep{chandrasekhar1957}, the magnetic field $\BB$ can be decomposed in spherical coordinates, into poloidal $\Phi(\xx)$ and toroidal
$\Psi(\xx)$ scalar functions, as explained in \Sec{subsec: spherical coordinates}. The initial magnetic field structure can be constructed by selecting a set of spherical harmonics (see \Eq{eq: phi and Psi scalar functions}), which defines the angular part of the magnetic field configuration. For additional details, consult Appendix~B of \citet{dehman2023a}.

As mentioned at the beginning of \Sec{sec: initial conditions}, our goal is to start the simulations with a locally isotropic spectrum characterized by an $\ell^3$ slope\footnote{The plotted energy spectrum (\Eq{eq: energy spectrum MATINS}) in \texttt{MATINS} is a 2D spectrum decomposed into spherical harmonics along $\theta$ and $\phi$ coordinates.}. Thus, we express the poloidal scalar function $\Phi(\xx)$ as:
\begin{equation}
    \Phi(\xx) = \frac{1}{r} \sum_{\ell m} \phi_{\ell m}(r_a) \,f_{\ell}(r) \, Y_{\ell m}(\theta,\phi),
\end{equation}
with 
\begin{equation}
   \Phi_{\ell m}(r_a) = \sum_{\ell m} \frac{1}{\ell(\ell+1)} \int dS_r \, B^r(r_a,\theta,\phi) \, Y_{\ell m}(\theta,\phi). 
\end{equation}
Here, $\Phi_{\ell m}(r_a)$ represents the weights of the multipoles, extracted from simulations using the \textsc{Pencil Code} at a given radial layer
$r_a$ beneath the surface of the star $R$. The surface differential is denoted by $dS_r$, $f_{\ell}(r)$ is the radial spectral mode function, and $B^r$ is the contravariant radial component of the magnetic field, as \texttt{MATINS} employs a nonorthogonal cubed-sphere metric. This selection ensures the achievement of the desired locally isotropic $\ell^3$ initial sub-inertial range.

To establish a locally isotropic magnetic field across all radial layers, we define $f_{\ell}(r)$ as follows:
\begin{equation}    
    f_{\ell}(r) = \sin (k^\mathrm{eff}_\ell r),\quad
    k^\mathrm{eff}_\ell =  \iota \sqrt{\ell(\ell+1)}/R,
    \label{eq: radial function MATINS}
\end{equation}
where $k^\mathrm{eff}_\ell$ is the radial wavenumber.
The choice of $k^\mathrm{eff}_\ell$ is motivated by the dimensional argument of the realizability condition;
see \Eqs{eq: realisability cond}{eq: realisability l}.
The parameter $\iota$ is a constant that can be adjusted in our simulations to ensure the desired number of multipoles along the radial direction, as one can define $k_0 R \propto \iota \, \ell_0$.
Here, $k_0$ and $\ell_0$ are the peak wavenumber and degree of the multipoles in the spectrum, respectively. For $\iota=1$, $k^\mathrm{eff}_\ell$ simplifies to the wavenumber $k$. 
We refer to \Sec{sec: different coordinates} for details regarding the selection of the $\iota$ parameter.

The choice of the radial spectral mode function $f_{l}(r)$, as defined, does not adhere to the potential and perfect conductor boundary conditions imposed at the surface and the crust-core boundary, respectively. However, considering our goal of conducting simulations with a large Lundquist number on the order of $100...1000$, it is anticipated that the initial condition's influence on the global evolution will be minimal. Also, as discussed above, the magnetic field is expected to readjust to the imposed boundary conditions after a few evolution time steps,
a behavior also observed in the \textsc{Pencil Code} simulations. It is worth noting that, in order to construct an initial field that respects boundary conditions, the solution of Bessel functions would be necessary, which inspired our current choice of the sine function.

To construct an initial helical field, the toroidal scalar function $\Psi(\xx)$ must be expressed as
\begin{equation}
        \Psi(\xx) =  k \, \Phi(\xx),
        \label{eq: Phi = k Phi}
\end{equation}
where $k= \sqrt{\ell(\ell+1)}/R$. Upon examining \EEq{eq: realisability l}, it becomes apparent that, at the surface of the star, the ratio of the left-hand side to the right-hand
side approaches one, indicating that the magnetic field tends toward maximal helicity. However, the presence of the $\Phi_{\ell m}^{\prime\;2}$ term on the right-hand side of \EEq{eq: realisability l} prevents the Chandrasekhar-Kendall formalism from achieving a fully helical field. It is worth noting that different choices of the radial function also ensure a helical field if \EEq{eq: Phi = k Phi} is enforced.

The magnetic field components are computed using the finite-volume curl operator applied to $\Phi(\xx)$ and $\Psi(\xx)$ in cubed-sphere coordinates; see \citet{dehman2023a}. This method guarantees that the initial magnetic field structure is divergence-free up to machine precision and effectively avoids the axis-singularity problem inherent in spherical coordinate. Our experience shows that this approach yields satisfactory results for our objectives, particularly in generating a visually more locally isotropic field, as will be demonstrated in \Sec{sec: Simulations}.

\subsection{Boundary conditions}
\label{sec: boundary conditions}
In our simulations, we confine the magnetic field to the crust of the star. Consequently, the inner boundary conditions are imposed by requiring the normal (radial) component of the magnetic field to vanish at the lower boundary ($r=r_0$). This physically represents the transition from normal to superconducting matter. Under these assumptions, the Poynting flux at the inner boundary is zero, preventing any energy flow into or from the core of the star.
In the \textsc{Pencil Code}, this boundary condition, in terms of the magnetic vector potential,
translates to
\begin{equation}
    \frac{\partial A_r}{\partial r} = A_\theta = A_\phi = 0 \quad (r=r_0).
\end{equation}
On the other hand, \texttt{MATINS} enforces vanishing inner boundary conditions for the radial component of the magnetic field ($B^r = 0$). This condition arises from the direct evolution of the magnetic field components within the code. Furthermore, the tangential components of the electric and magnetic fields in cubed-sphere coordinates at the inner boundary are specified as follows:
\begin{eqnarray}
&& E^{\xi, \eta}(r_0)= \frac{1}{2}  E^{\xi, \eta}(r_0+dr),\nonumber\\
&& B^{\xi, \eta}(r_0-dr) = \frac{r_0}{r_0-dr} B^{\xi, \eta}(r_0).
  \label{eq: odd-even decoupling solution inner BC}
\end{eqnarray}
In the equations above, we have omitted the angular coordinates for brevity. The radial coupling among the nearest neighbors occurring over a distance $dr$ (see \Eq{eq: odd-even decoupling solution inner BC}), is implemented to address the issue of odd-even decoupling or checkerboard oscillations, which arise from two slightly different solutions—one corresponding to odd grid points and the other to even grid points. 
This phenomenon is a known issue with second-order central difference schemes when applied to the second derivative of a function. By adopting this approach, we mitigate tangential currents at the crust-core interface and improve stability during the evolution.

At the outer boundary, all components of the magnetic field are continuous, if surface current sheets are excluded. In the \textsc{Pencil Code} simulations, the magnetic field is radial
at the outer boundary, which, in terms of the magnetic vector potential, translate to
\begin{equation}
    A_r = 0, ~~ \frac{\partial A_\theta}{\partial r} = - \frac{A_\theta}{r}, ~~ \frac{\partial A_\phi}{\partial r} = - \frac{A_\phi}{r} ~~ (r = R). 
\end{equation}
Along the $\theta$ (or $y$) direction, perfect boundary conditions were imposed.  
Moreover, in the \textsc{Pencil Code}, we sometimes use periodic boundary conditions (\Sec{sec: Periodic BC}).

Conversely, in \texttt{MATINS}, we implement an external potential (current-free) solution for the magnetic field at the surface of the star, governed by $\bnabla \times \BB = 0$ and
$\bnabla \cdot \BB=0$. This magnetic field is expressed as the gradient of a magneto-static potential that satisfies the Laplace equation. The potential is expanded in spherical harmonics, allowing us to express the three components of the magnetic field at the surface of the star as
 \begin{equation}
        B^r = B_0 \sum_{\ell m} (\ell+1) \, b^m_\ell  Y_{\ell m}, 
     \label{eq: spectral decomposition of Br}  
 \end{equation}
where $B_0$ is a normalization factor and $b^m_{\ell}$ corresponds to the dimensionless weight of the multipoles:
 \begin{equation}
     b^m_\ell = \frac{1}{B_0(\ell+1)} \int \frac{dS^r}{r^2} B^r Y_{\ell m}.
 \end{equation}
The angular components of the magnetic field are given by:
 \begin{equation}
    B^{\theta} = - B_0 \sum_{\ell m}  b^m_\ell  \frac{\partial Y_{\ell m}}{\partial \theta},
 \end{equation}
 \begin{equation}
    B^{\phi} = - \frac{B_0}{\sin\theta} \sum_{\ell m} b^m_\ell \frac{\partial Y_{\ell m}}{\partial \phi}.
\end{equation}
The angular components of the magnetic field ($B^\theta$ and $B^\phi$) are then converted into the $B^\xi$ and $B^\eta$ components on the
cubed-sphere grid within the code. For a detailed description of the magnetic boundary condition in \texttt{MATINS}, see \citet{dehman2023a}.

\subsection{Units and simulation parameters}
\label{sec: units and simulations parameters}

For the \textsc{Pencil Code} simulations, we present the results in adimensional form by introducing the following units
\begin{equation}
    [x] = R, \quad [t] = R^2/ \eta , \quad [\BB]= e n_e \mu_0 \eta. 
\end{equation}
This implies that the current density is measured in units of $[\JJ] = [\BB]/ \mu_0 R$. To express the \textsc{Pencil Code} results in dimensional units, we multiply by the appropriate units described above. Moreover, we consider a Hall cascade with a characteristic wavenumber $k_0$, which represents the initial peak of the spectrum. This peak wavenumber is related to the spherical harmonic degree $\ell_0$,
where most of the energy is concentrated, through the relation 
$k_0 = \sqrt{\ell_0(\ell_0+1)}/R$.
We set $1/e n_e \mu_0 = 1$ 
and adopt a time-dependent prescription for the magnetic diffusivity $\eta$:
\begin{equation}
  \eta(t) =
    \begin{cases}
       2\times 10^{-4} & \text{if $t\leq 0.1$}\\
      2\times 10^{-4} \, \mathrm{exp}\left(-0.4 \, t\right) & \text{if $t>0.1$}
    \end{cases}     
    \label{eq: eta pencil}
\end{equation}
Here, $\eta$ remains constant throughout most of the evolution time, except for the very late stage (after $t > 0.1$). 
This adjustment ensures an increase in the Lundquist number at later times ($t > 0.1$) if the simulation continues for that duration. 
We ignore in this study the depth-dependence of the magnetic diffusivity $\eta$ and the electron number density $n_e$.
Additionally, for the \textsc{Pencil Code}, we use different mesh points depending on the aspect ratio considered in each simulation. The specific mesh points used are detailed in \Tab{Ttimescale}.

In \texttt{MATINS}, we use physical units commonly applied in astrophysics, defined as follows
\begin{eqnarray}
\hspace{40pt}
 [x] = 1\,\text{km}, \quad [t] = 1\,\text{Myr}, \quad [\BB] = 1\,\text{G}, &&\nonumber\\
\quad [\JJ] = 1\,\text{G Myr}^{-1},\quad  [\EM(\ell)] = 1\,\text{erg}.  &&
\end{eqnarray}
We have set the average magnetic field strength to approximately $10^{12}$\,G
and fixed the Hall prefactor at $1/e n_e \mu_0 = 1\km^2\Myr^{-1}(10^{12}\G)^{-1}$.
Instead of varying these parameters, we adjusted the magnetic diffusivity
$\eta$ (as shown in \Tab{table: MATINS runs}), ignoring the depth-dependence of the electric conductivity, to achieve a Lundquist
number $\Lu$ of order 100 (\Eq{eq: Lu}), consistent with the magnetic field strengths
typical of magnetars ($10^{14}\dots 10^{15}$\,G).
Additionally, the grid consists of $64 \times 47^2 \times 6$ mesh points,
with $n_r = 64$ radial mesh points, $n_\xi = n_\eta = 47$ angular mesh
points, and 6 patches forming the cubed-sphere structure.

In the following section, we will use $t$ (in Myr) to represent
time in \texttt{MATINS} simulations and the dimensionless
$\tilde{t} = \eta t/R^2$ for time in the \textsc{Pencil Code} simulations
(see also \Tab{tab: time run R1} for the correspondence of our times
in various units in the \textsc{Pencil Code} simulations).
To compare time between the two codes, it is essential to use a dimensionless quantity.
Specifically, we compare the value of $\eta t/R^2$.

\section{Simulations}
\label{sec: Simulations}

\begin{table*}[htp]
\caption{Summary of the \textsc{Pencil Code} simulations.}
\centering
\vspace{-5pt}
\begin{tabular}{lccccccccccc}
Run & geom/B.C. & $r/R$ & $\theta/\pi$ & $\phi/\pi$ &  $\ell_0(\Tilde{t}_0)$ & $\ell_0(\Tilde{t}_2)$ & $\Lu(\Tilde{t}_1)$ & $\Lu(\Tilde{t}_2)$ & $k_0\xi (\Tilde{t}_0)$ & $\chi(\Tilde{t}_0)$ & Mesh points\\
\hline
R1 & Cart/PC-VF & 0.9...1 &   0...1   & 0...1   & 200 &  50 &260 & 132 & 0.50 & 1.19 &  $64\times 2048^2$ \\
R2 & Sph/PC-VF  & 0.9...1 & 0.1...0.9 & 0...1   & 200 &  60 &251 & 124 & 0.60 & 1.12 &  $64\times 2048^2$ \\
R3 & Cart/PC-VF & 0.8...1 &   0...1   & 0...1   & 200 &  30 &361 & 202 & 0.61 & 1.16 & $128\times 2048^2$ \\
R4 & Cart/PC-VF & 0.8...1 &   0...0.5 & 0...0.5 & 200 &  10 &487 & 272 & 0.61 & 1.15 & $128\times 1024^2$ \\
R5 & Cart/PC-VF & 0.8...1 &   0...0.5 & 0...0.5 & 50  &  10 &724 & 464 & 0.46 & 0.87 & $128\times 1028^2$ \\
R6 & Cart/P     & 0.9...1 &   0...1   & 0...1   & 200 &  60 &283 & 168 & 0.61 & 1.19 &  $64\times 2048^2$ \\
\end{tabular}
\tablefoot{Cartesian simulations are indicated by `Cart', while spherical simulations
are denoted by `Sph'. Simulations with perfect conductor boundary conditions on the inner surface and vertical field boundary conditions on the outer surface in the
radial directions are labeled as `PC-VF'. By contrast, simulations with periodic boundary conditions in the radial directions are labeled as `P'.
Data is shown at specific times:
$\Tilde{t}_1=\eta t_1 / R^2 = 2 \times 10^{-6}$ and $\Tilde{t}_2=\eta t_2 / R^2 = 2 \times 10^{-5}$. We note that $\ell_0 = k_0 R$ for the Cartesian runs. }
\label{Ttimescale}
\end{table*}

Constructing the NS background model entails solving the Tolman-Oppenheimer-Volkoff equations \citep{oppenheimer1939}, considering a nuclear EoS at zero-temperature\footnote{NSs consist of degenerate matter, typically characterized by temperatures lower than the Fermi temperature for their entire existence. In this specific temperature range, quantum effects, as dictated by Fermi statistics, overwhelmingly dominate over thermal effects. Therefore, the EoS for NSs can be effectively approximated as that of zero temperature, allowing us to largely ignore thermal contributions for most of their lifespan.}. This approach involves describing both the liquid core and the solid crust of the star. Through these calculations, we can determine the thickness of the NS crust, estimated to be approximately $1$\,km in relation to its overall radius of about $10$\,km.
This estimation reveals the extreme aspect ratio of the crust,
approximately 1:30, with $\mathcal{A}=(R-r_0)/ \pi R$.
In our simulations conducted in Cartesian, spherical (\textsc{Pencil Code}), and cubed-sphere (\texttt{MATINS}) coordinates, we account for the actual aspect ratio of the NS crust.

Below, we delve into the various runs relevant to our study. We focus on simulations favoring the inverse cascade in the crust of a NS.
The simulations conducted with the \textsc{Pencil Code} are detailed in \Tab{Ttimescale}, while those performed using \texttt{MATINS} are outlined in \Tab{table: MATINS runs}.

\subsection{Reference run}
\label{sec: ref run}

In this section, we describe the reference run in our study, run R1, performed using the \textsc{Pencil Code} in Cartesian coordinates (see \Tab{Ttimescale}). This run is constructed based on our experience and represents the optimal configuration to validate the inverse cascade in a NS crust.
Run~R1 is characterized by an average value of $\Lu$ on the order of a few hundred. 
In our Cartesian domain, the crustal shell extends from $x=0.9 ... 1$, $y = 0 ... 1$, and $z = 0 ... 1$,
corresponding to spherical coordinates in the range $r/R=0.9 ... 1$, $\theta/\pi = 0 ... 1$, and $\phi/\pi = 0 ... 1$. 

The initially helical magnetic field for run~R1 is formulated as detailed
in \Sec{sec: initial conditions}.
Given the extreme aspect ratio ($\mathcal{A}\approx$ 1:30) of the NS
crust, attributed to its small thickness, our experience has shown that
a sufficiently small-scale magnetic field structure is necessary to
facilitate inverse cascading.
As a result, the magnetic spectrum should predominantly feature
small-scale structures with $\ell_0  = k_0 R \approx 200$.
This corresponds to a wavelength $2\pi/k_0=0.03\,R$, which is about
one-hundredth of the latitudinal extent but only about one third of the
crust's depth.

\begin{figure}[!htbp]
\begin{center}
\hspace{-0.5cm}
\includegraphics[width=\columnwidth]{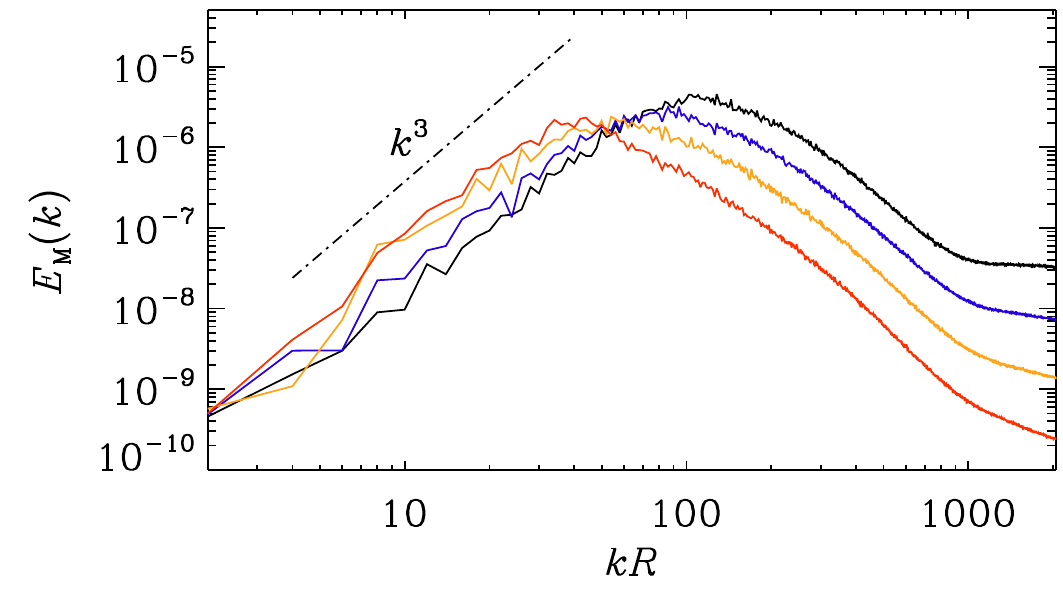}
\end{center}
\caption{Spectral energy of the Cartesian reference run~R1. The magnetic spectra are displayed at $\Tilde{t}=2 \times 10^{-6}$ (black), $\Tilde{t}=6 \times 10^{-6}$ (blue), $\Tilde{t}=2 \times 10^{-5}$ (yellow),
and $\Tilde{t}=6 \times 10^{-5}$ (red). }
\label{pspectra}
\end{figure}

In Figure~\ref{pspectra}, we present magnetic energy spectra of run~R1 at
four distinct dimensionless times: $\Tilde{t}=2 \times 10^{-6}$ (black),
$6 \times 10^{-6}$ (blue), $2 \times 10^{-5}$ (yellow), and $6 \times 10^{-5}$ (red).
These times correspond to the normalized times $t/\tau_\mathrm{Ohm} = 0.11$, $0.15$, $0.20$, and $0.26$, respectively, where the Ohmic diffusion timescale is defined as $\tau_\mathrm{Ohm} = \xi^2/\eta$
(see \Tab{tab: time run R1}). 
The spectra reveal an inverse cascade of the magnetic peak evolving underneath $k^{3/2}$ envelope (not shown). 
For times $\Tilde{t}> 2\times 10^{-5}$, the peak of the magnetic energy spectrum no longer decays over time (yellow and red curves in Figure~\ref{pspectra}).
As explained at the end of \Sec{subsec: cart coord}, this indicates that the magnetic field has reached a maximally helical state, while the total magnetic energy continues to dissipate.

\begin{table}
\centering
\caption{Correspondence of times in various units for the \textsc{Pencil Code} simulations.}
    \begin{tabular}{cccc}
         $t$ [$R^2/\eta$] & $\tilde{t}\equiv \eta t/R^2 $& $\widetilde{t}\equiv \eta t /(R-r_0)^2$ &$t / \tau_\mathrm{Ohm}$ \\
         \hline
         0.01  & $2 \times 10^{-6}$ & $2 \times 10^{-4}$ & 0.11\\
         0.03  & $6 \times 10^{-6}$ & $6 \times 10^{-4}$ & 0.15\\
         0.1   & $2 \times 10^{-5}$ & $2 \times 10^{-3}$ & 0.20\\
         0.3   & $6 \times 10^{-5}$ & $6 \times 10^{-3}$ & 0.26\\
    \end{tabular}
    \tablefoot{$t$ represents the time in \textsc{Pencil Code} units, $\tilde{t}$ shows the dimensionless time normalized by the stellar radius, $\widetilde{t}$ is the time normalized by the crust thickness,
and $t/\tau_\mathrm{Ohm} \equiv \eta t / \xi^2$ provides the normalized time with respect to the Ohmic diffusion timescale, $\tau_\mathrm{Ohm} = \xi^2 / \eta$.}
    \label{tab: time run R1}
\end{table}

Toward the end of the simulation, we notice that the inverse cascade
stalls when $\ell_0 \approx 30$, with almost no further transfer of energy toward
the dipolar component.
In the following sections, we examine the impact of the
coordinate system, peak wavenumber position, 
aspect ratio, and boundary conditions to understand
what is limiting the inverse cascade from transferring energy to
larger-scale structures.

\subsection{Different coordinate systems}
\label{sec: different coordinates}
To study the impact of the coordinate systems on the inverse
cascade, we replicated run R1 in spherical coordinates using
\textsc{Pencil Code} in run R2 (see \Tab{Ttimescale}) and in
cubed-sphere coordinates using \texttt{MATINS} in run R$\iota5$
(see \Tab{table: MATINS runs}).
These two runs have an average value of $\Lu$ similar to that of run R1,
both on the order of a few hundred.
The crustal shell is in the range $r/R=0.9 ... 1$ for both runs.
In run R2, we choose $\theta/\pi = 0.1 ... 0.9$ to avoid the axis
singularity problem in 3D spherical coordinates, while $\phi/\pi = 0 ... 1$.
By contrast, run R$\iota5$ covers the full crustal shell and has
$\theta/\pi = 0 ... 1$, and $\phi/\pi = 0 ... 2$.

\begin{figure}[!htbp]
\includegraphics[width=\columnwidth]{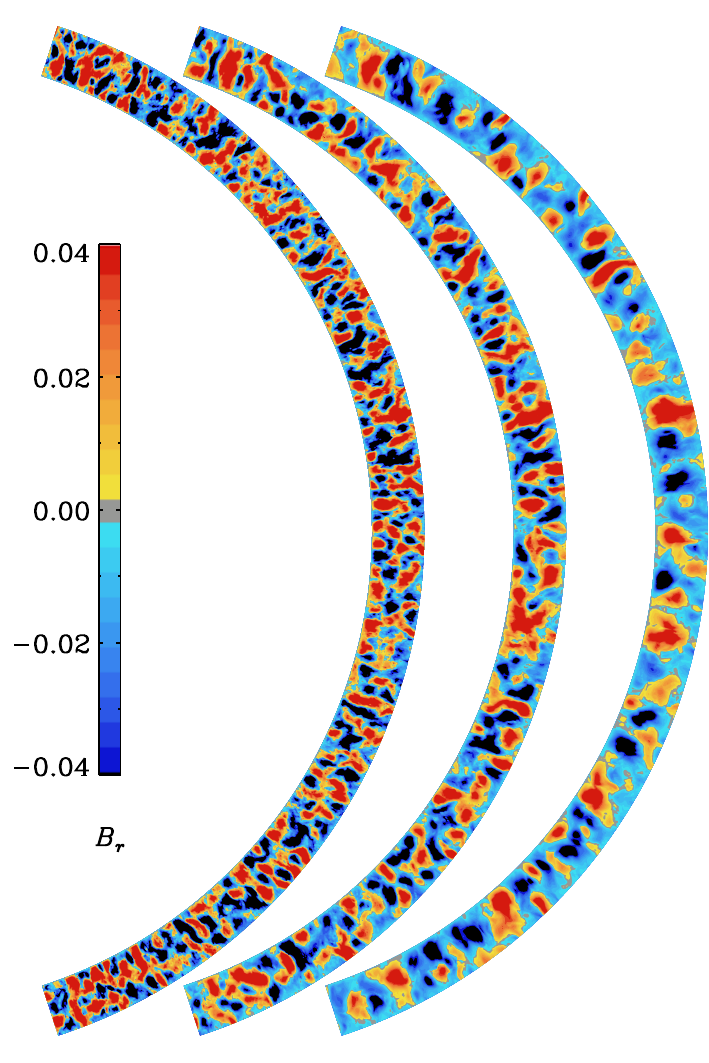}
\caption{Meridional slices of $B^r(r,\theta)$ for run R2, with
$\ell_0(\Tilde{t}_0)=200$ at $\Tilde{t} = 6 \times 10^{-7}$, $2 \times 10^{-6}$, and $6 \times 10^{-6}$ (from left to right).
}
\label{eps: pslice_rt_64x2048x2048t8_0p4_k200}
\end{figure}

\begin{figure}[!htbp]
\begin{center}
\hspace{-1cm}
\includegraphics[width=\columnwidth]{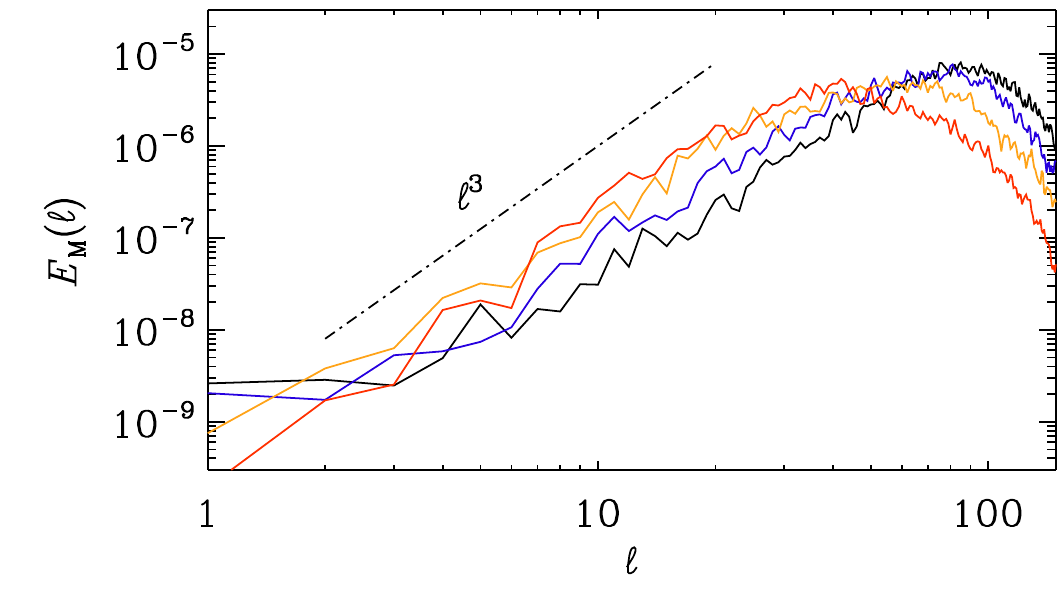}
\end{center}
\caption{Spectral energy of the spherical run~R2. As in Figure~\ref{pspectra}, the magnetic spectra are shown at 
$\Tilde{t}= 2 \times 10^{-6}$ (black), $ 6 \times 10^{-6}$ (blue), $ 2 \times 10^{-5}$ (yellow) and $6 \times 10^{-5}$ (red). 
}
\label{pspectra R2}
\end{figure}

The meridional slices of the $B^r(r,\theta)$ component of the magnetic field for run R2 are illustrated in Figure~\ref{eps: pslice_rt_64x2048x2048t8_0p4_k200} at three
different times: $\Tilde{t}= 6 \times 10^{-7}$, $2 \times 10^{-6}$, and $6 \times 10^{-6}$ (from left to right).
Initially, the magnetic spectrum predominantly features small-scale structures on the order of $k_0 R \approx 200$ (at $\Tilde{t}= 6 \times 10^{-7}$).
As the evolution processes, the emergence of large-scale structures from small-scale ones becomes increasingly apparent,
ultimately resulting in magnetic field structures comparable in size to the NS crust.
This growth in scale hints at why the inverse cascade concludes, which can be attributed to the extreme aspect ratio $\mathcal{A}$ of the NS crust.
The inverse cascade phenomenon observed in the magnetic field of run R2 is further corroborated by the magnetic energy spectra shown in Figure~\ref{pspectra R2} at various times.

To verify if the same behavior occurs with cubed-sphere coordinates, we conducted simulations using \texttt{MATINS}. This test is also essential for determining if two different numerical codes yield similar results. As demonstrated previously using the \textsc{Pencil Code}, for the inverse cascade to occur, the initial magnetic spectrum should predominantly exhibit small-scale structures with $\ell_0\approx 200$. However, achieving the necessary resolution with \texttt{MATINS} is challenging because it is not parallelized, unlike the \textsc{Pencil Code}. 

\begin{table*}[!htbp]
\caption{Summary of \texttt{MATINS} simulations on a cubed-sphere grid representing a crustal shell.}
\centering
\begin{tabular}{lccccccccccc}
Run &  $r/R$ & $\theta/\pi$ & $\phi/\pi$ &   $\ell_0$ &
$\eta$ [km$^2$/Myr] & $\Lu(t_0)$ & $\Lu(t_3)$ & $k_0 \xi (t_0)$ & $\chi (t_0) $ & Mesh points\\
\hline
R$\iota1$   & 0.9...1.0 & 0...1 & 0...2 & 40 & 
$8   \times 10^{-3}$ & 60  & 58 & 0.54  & 0.72 &$64\times47^2\times6$ \\ 
R$\iota3$   & 0.9...1.0 & 0...1 & 0...2 & 120 &
$4.5 \times 10^{-3}$ & 103 & 98 & 0.63 & 0.85 &$64\times47^2\times6$   \\
R$\iota5$   & 0.9...1.0 & 0...1 & 0...2 & 200 &
$1.2 \times 10^{-2}$ & 107 & 90 & 0.64 & 0.92 &$64\times47^2\times6$ \\
 R$\iota10$  & 0.9...1.0 & 0...1 & 0...2 & 400 & 
 $1.5 \times 10^{-2}$ & 106 & 91 &   0.40  & 0.69 & $64\times47^2\times6$ 
\end{tabular}
\tablefoot{Data is shown at specific, fixed times: $t_0 = 0.0$, $t_1=0.005$, $t_2=0.01$ and $t_3 = 0.02$\,Myr, resulting in $\eta t_1/R^2 = 6 \times 10^{-6}$ and $\eta t_3/R^2 = 2 \times 10^{-5}$, with $\eta$ set to $1.2 \times 10^{-2}$\,km$^2$/Myr. We note that in \texttt{MATINS}, $\ell_0 \equiv \iota \ell_0^\mathrm{afford}$. }
\label{table: MATINS runs}
\end{table*}

An alternative approach involves concentrating more structures exclusively
in the radial direction.
This initial magnetic field configuration can be conceptualized as
a squashed magnetic field structure, achieved by selecting an $\iota$
value greater than one; see \Eq{eq: radial function MATINS}.
The $\iota$ parameter is determined based on the results from
run R1.
During run R1, the initial value $\ell_0 (\Tilde{t}_0) \approx 200$
decreased to $\ell_0 (\Tilde{t}= 6 \times 10^{-5}) \approx 30$.
By this time, the radial scale has reduced to $\pi/k\approx0.1$, corresponding to the crustal thickness, and the inverse cascade concluded.
This indicates that for the inverse cascade to remain active, $\ell$ must not fall significantly below the inverse aspect
ratio $\gtrsim \mathcal{A}^{-1}$.
To span a factor $\mathcal{F}\approx200/30$ in
the inverse cascade during \texttt{MATINS} simulations with
$\ell_\mathrm{max} = 80$, where the affordable spherical degree at the peak is
$\ell_0^\mathrm{afford}\approx40$, the radial scale must be reduced by a squashing factor $\iota$.
The value of $\iota$ can be computed using the following equation:
\begin{equation}
\iota \gtrsim \frac{\mathcal{F}}{ \mathcal{A} \, \ell_0^\mathrm{afford}}.
\label{eq: iota}
\end{equation}
For example, with $\ell_0^\mathrm{afford} \approx 40$ in \texttt{MATINS},
an $\iota$ value of approximately 5 is required to accommodate
$\ell_0 \approx 200$.

\begin{figure}[!htbp]
\begin{center}
\includegraphics[width=\columnwidth]{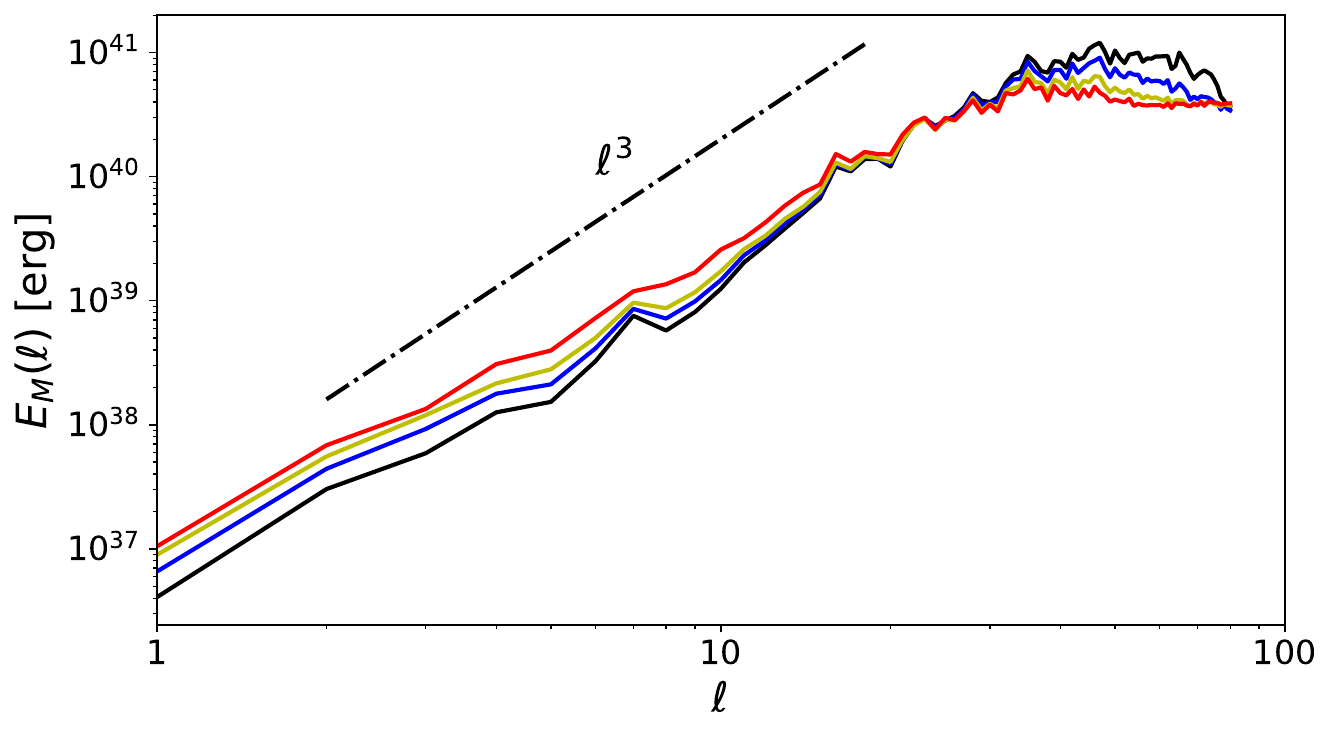}
\end{center}\caption{Spectral energy of the cubed-sphere run R$\iota 5$ at $t_0=0.0$
(black), $t_1=0.005$\,Myr (blue), $t_2=0.01$\,Myr (yellow), and $t_3 = 0.02$\,Myr (red).}
\label{fig: cubed-sphere iota5}
\end{figure}

\begin{figure}[!htbp]
\includegraphics[width=\columnwidth]{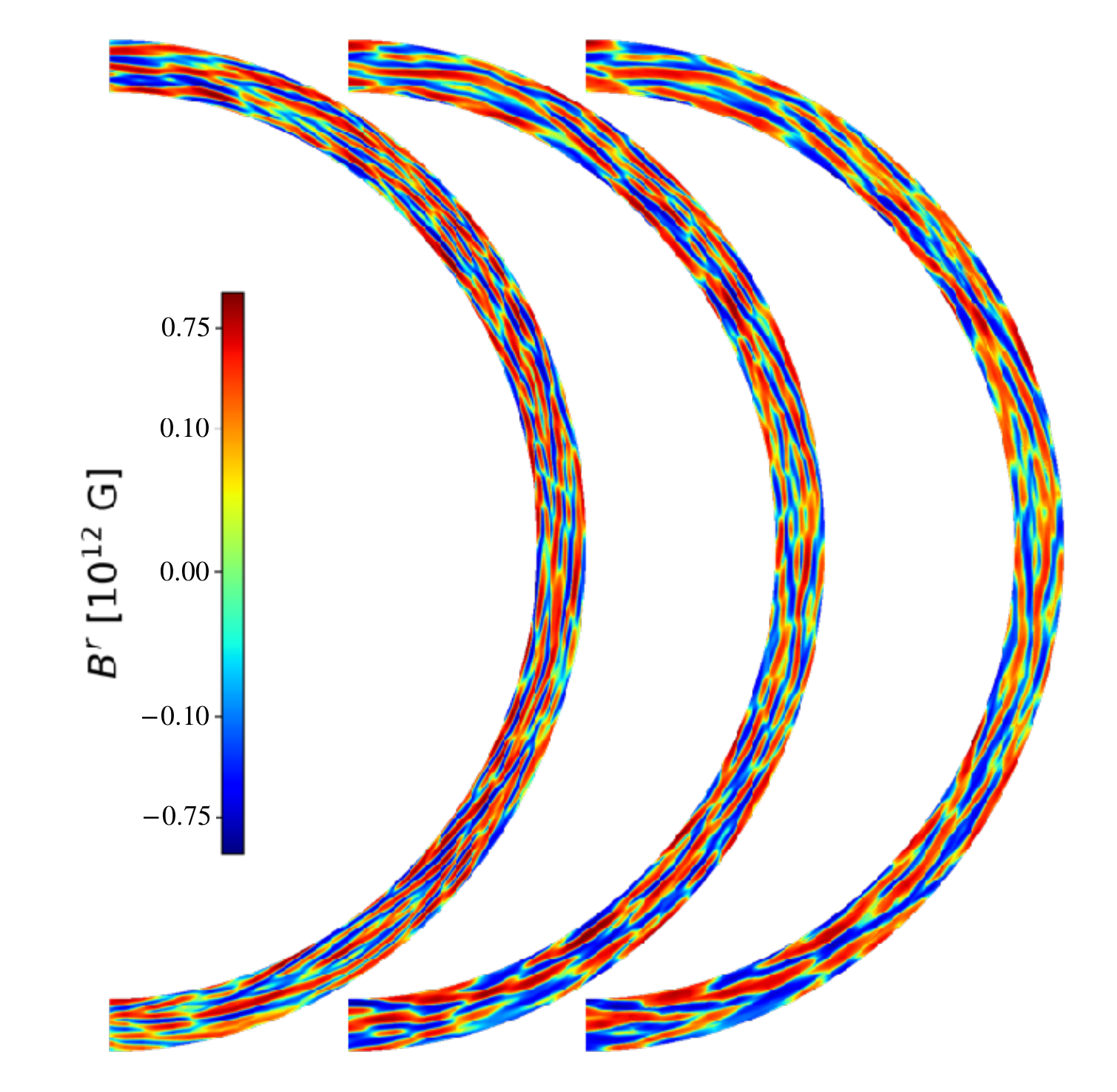}
\caption{Meridional slices of $B^r(r,\theta)$ for run R$\iota$5 at $t=0.0$, $0.01$\,Myr and, $0.02$\,Myr (from left to right).}
\label{fig: Riota5 contour}
\end{figure}

Figure~\ref{fig: cubed-sphere iota5} presents the magnetic energy spectra for the R$\iota5$ simulation at four different times: $t_0=0.0$ (black), $t_1=0.005$ (blue), $t_2=0.01$ (yellow), and $t_3=0.02$\,Myr (red). 
The Ohmic diffusion timescale, defined as $\xi^2/\eta$, is approximately 0.02 Myr.
Similar to the Cartesian and spherical runs (R1 and R2) conducted with the \textsc{Pencil Code}, the magnetic field in run R$\iota5$ exhibits an inverse cascade, featuring ascending $\ell^3$ spectra and demonstrating temporal decay that scales as $\ell^{3/2}$ on the cubed-sphere grid (not shown). 
Figure~\ref{fig: Riota5 contour} shows meridional slices of the $B^r(r,\theta)$ component of the magnetic field for run R$\iota5$ at $t=0.0$, $0.01$ and, $0.02$\,Myr. The radial magnetic field structures are visibly squashed along the radial direction. Notably, the inverse cascade observed in \texttt{MATINS} is less pronounced than in the \textsc{Pencil Code}, as demonstrated in Figures~\ref{fig: cubed-sphere iota5} and \ref{fig: Riota5 contour}. This discrepancy may be related to the $\iota$ parameter (\Eq{eq: iota}), which represents the squashing factor.

The consistent behavior observed in run R1, run R2, and run R$\iota5$, which utilize Cartesian, spherical, and cubed-sphere coordinates respectively, indicates that the choice of coordinate system is not critical as long as the initial conditions (and the correct aspect ratio) are nearly identical.
Notably, despite the differing geometrical coordinates, all three models exhibit an inverse cascade limited to $\ell_0 \approx 30$, with minimal energy transfer to the dipolar component.

\subsection{Effects of peak position}
\label{sec: effect of peak position}
In this section, we examine the effect of varying initial $k_0 R$ values on the inverse cascade using \texttt{MATINS} simulations. In the first case, we set $\iota = 1$ (see \EEq{eq: iota}) for run R$\iota$1, resulting in $k_0 R = 40$. In the second case, $\iota$ was increased to 10 for run R$\iota$10, resulting in $k_0 R = 400$. These simulations were designed to test our interpretation of the squashed initial field configuration in \texttt{MATINS} and evaluate whether it yields the anticipated results.

\begin{figure}[t]
\begin{center}
\includegraphics[width=0.9\columnwidth]{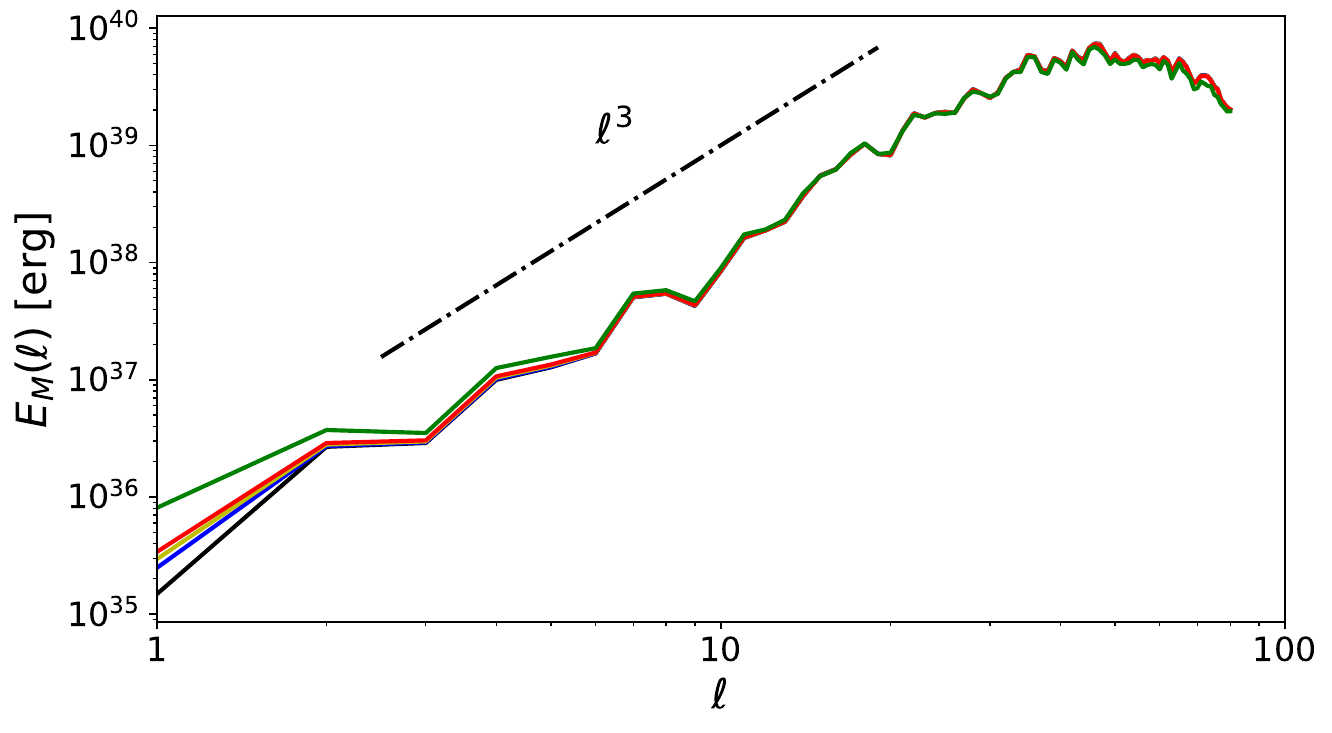}
\includegraphics[width=0.9\columnwidth]{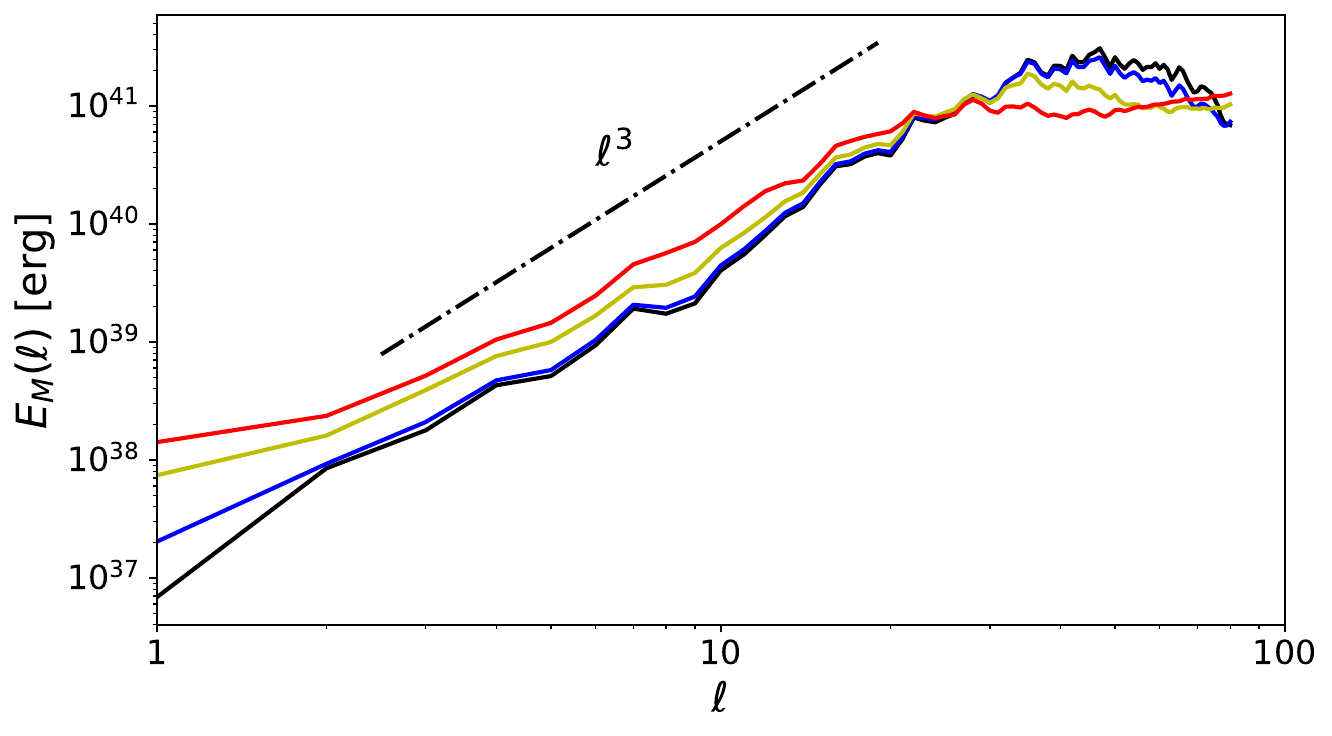}
\end{center}\caption{Spectral magnetic energy for run R$\iota1$ (upper panel) and run R$\iota10$ (lower panel). As in Figure~\ref{fig: cubed-sphere iota5}, the magnetic spectra are shown at $t=0.0$ (black), $0.005$ (blue), $0.01$ (yellow), $0.02$ (red), and $0.04$\,Myr (green).}
\label{fig: cubed-sphere runs smaller and larger kR.}
\end{figure}

Figure~\ref{fig: cubed-sphere runs smaller and larger kR.} presents the results for runs R$\iota$1 (upper panel) and R$\iota$10 (lower panel) at various stages of evolution. In run R$\iota$1, no inverse cascade is observed, likely because the field structures are comparable in size to the crustal thickness. In contrast, run R$\iota$10 exhibits a clear and more pronounced inverse cascade compared to run R$\iota$5.
These results suggest that the presence of small-scale initial spherical degrees is crucial for the inverse cascade to occur.

\subsection{Aspect ratio}
\label{sec: aspect ratio}
To study the impact of the aspect ratio on the inverse cascade, we conducted a replication of the reference run R1, and in this iteration, we adjusted solely the crust aspect ratio.
On the one hand, in the new run R3 (see \Tab{Ttimescale}),
we doubled the crustal thickness. Consequently, the crustal shell is now in the range $r/R=0.8 ... 1$,
accounting for one-fifth of the entire NS cross-sectional area, as opposed to the previous one-tenth.
On the other hand, in the new run R4, we also doubled the crustal thickness and, in addition, we reduced $\theta/\pi$ and $\phi/\pi$ from the range of $0 ... 1$ to $0 ... 0.5$. Therefore, the geometry of our domain now approaches a cubic configuration.

\begin{figure}[!htbp]
\begin{center}
\includegraphics[width=\columnwidth]{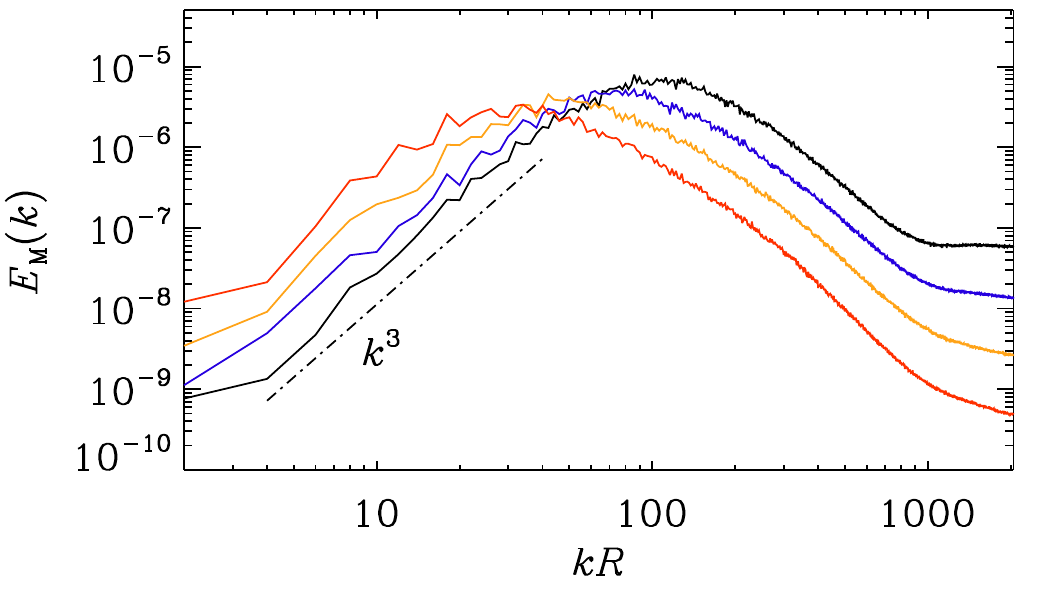} %
\includegraphics[width=\columnwidth]{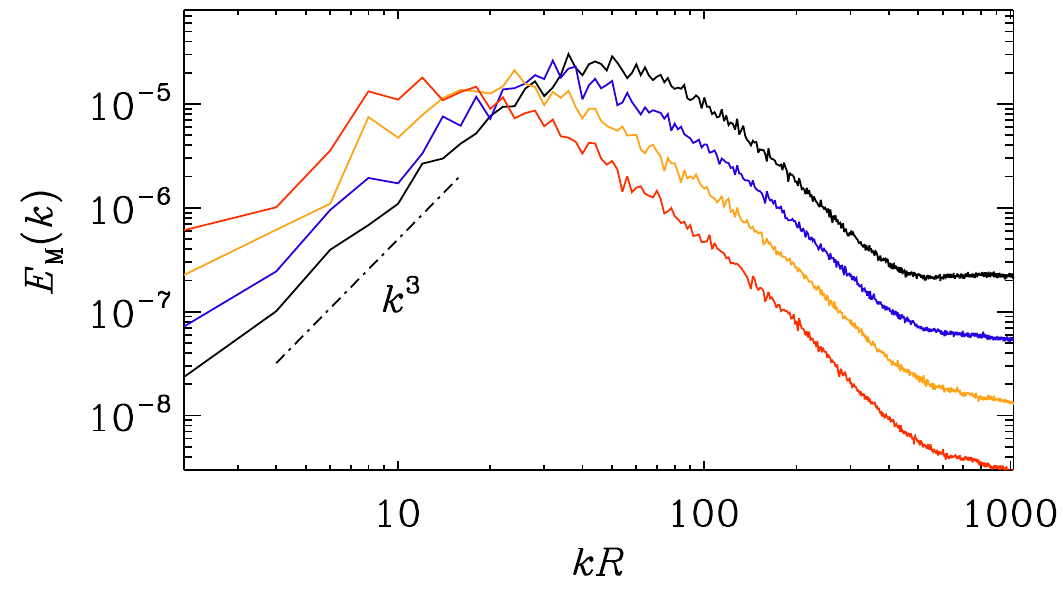} %
\end{center}\caption{Spectral energy of run R3 (upper panel) and run R4 (lower panel) in Cartesian coordinates, each characterized by different aspect ratios compared to the reference run R1. Data is shown at
$\Tilde{t}= 6 \times 10^{-7}$ (black), $ 2 \times 10^{-6}$ (blue), $ 6 \times 10^{-6}$ (yellow) and $2 \times 10^{-5}$ (red).
}
\label{fig: aspect ratio impact on peak}
\end{figure}

The magnetic energy spectra for runs~R3 and R4 are depicted in the upper and lower panels of Figure~\ref{fig: aspect ratio impact on peak}, respectively.
In this representation, magnetic spectra are plotted at 
$\Tilde{t}=6\times 10^{-7}$ (black), $2\times 10^{-6}$ (blue), $6\times 10^{-6}$ (yellow), and $2\times 10^{-5}$ (red). 
Compared to the reference run~R1 (Figure~\ref{pspectra}), R3 and R4 exhibit a more pronounced inverse cascading phenomenon with several distinct behaviors.

At $\Tilde{t} = 2 \times 10^{-5}$, the characteristic scale $\ell_0$ is approximately 20 for run~R3 and 10 for run~R4, whereas for run~R1, $\ell_0 \approx 50$ at the same time. 
Furthermore, at largest scale ($\ell \sim 1$), the spectral magnetic energy remains nearly constant in run~R1, with $\EM(k) \approx 6 \times 10^{-10}$. In contrast, runs~R3 and R4 show significant energy accumulation in the dipolar component ($\ell \sim 1$), with $\EM(k) \approx 10^{-8}$ for run~R3 and $\EM(k) \approx 10^{-6}$ for run~R4.

Compared to run~R3 and the reference run~R1, run~R4 demonstrates a more rapid inverse cascade, achieving a fully helical field state in a shorter time. For $\tilde{t} \geq 2\times 10^{-6}$, the peak of the magnetic energy spectrum, $\EM^\mathrm{max}$, in run~R4 is steadily maintained throughout its evolution, indicating a convergence toward a maximally helical field. Meanwhile, the total magnetic energy continues to dissipate.

These differences relative to the reference run R1 are attributed to the aspect ratio, particularly in run R4, where the geometry approximates cubic symmetry, while run R3 involves only doubling the crustal thickness. Variations in crustal geometry 
significantly influence the inverse cascade,
the efficiency of achieving a maximally helical field,
and the formation of the large-scale magnetic field.
The realistic extreme aspect ratio of the NS crust limits the inverse cascade, and the energy transfer toward the dipolar component.

At this stage, it is intriguing to explore whether a less extreme aspect ratio, as used in run R4, can reveal the presence or absence of an inverse cascade for a smaller initial $\ell_0$. To investigate this, we replicated run R4, adjusting $\ell_0$ from approximately 200 to 50 for run R5, as detailed in \Tab{Ttimescale}. The magnetic energy spectra for run R5
exhibit behavior similar to that of run R4, confirming the presence of an inverse cascade even with an initial peak wavenumber of $\ell_0 \approx 50$. 
Notably, at $\Tilde{t}_2= 2\times 10^{-5}$, the peak wavenumber for both runs R4 and R5 reaches $\ell_0 \approx 10$, as shown in \Tab{Ttimescale}. As previously mentioned, this outcome is influenced by the crustal geometry used in both runs, which closely resembles a cubic configuration. We note that starting with $\ell_0 \approx 50$ in a model replicating the reference run R1 does not result in an inverse cascade.

\subsection{Periodic boundary conditions}
\label{sec: Periodic BC}
In this section, we examine the impact of magnetic boundary conditions on our simulations.
To do so, we replicate reference run R1, but modify the magnetic boundary conditions by replacing the perfect conductor and vertical field boundary conditions on the inner and outer radial boundaries, respectively, with periodic boundary conditions.
In \Tab{Ttimescale} it is labeled as run~R6.

\begin{figure}[t]
\begin{center}
\hspace{-1cm}
\includegraphics[width=\columnwidth]{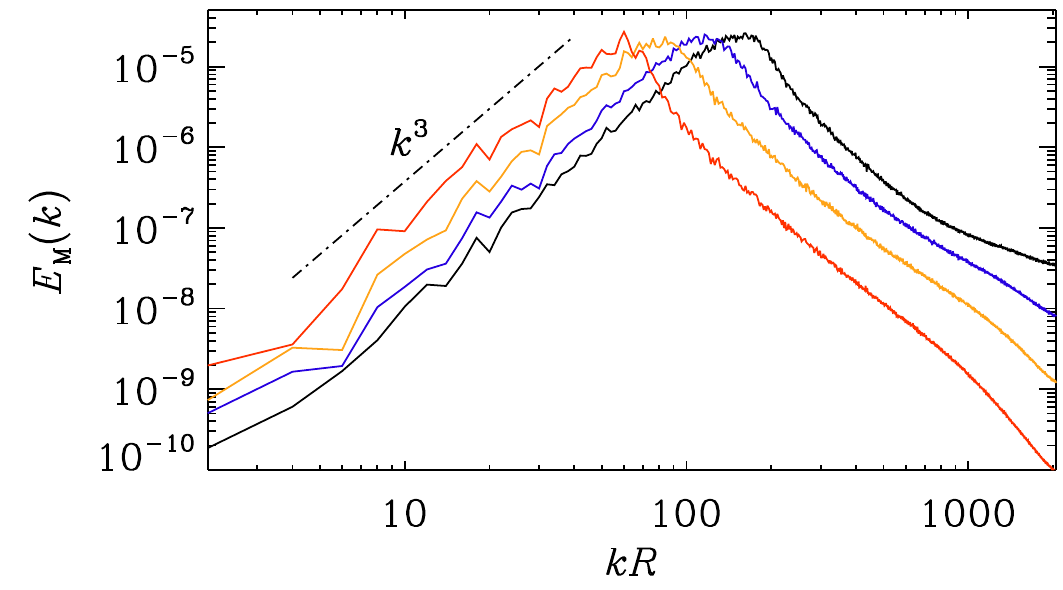} 
\end{center}
\caption{Spectral energy of run R6 in Cartesian geometry, characterized using periodic boundary conditions compared to reference run R1. Data is shown at $\Tilde{t}=6 \times 10^{-7}$ (black), $2\times 10^{-6}$ (blue), $6 \times 10^{-6}$ (yellow), and $2 \times 10^{-5}$ (red).
}
\label{fig: PC64x2048x2048t8_0p4_k200}
\end{figure}

The magnetic spectrum of run R6 is illustrated in Figure~\ref{fig: PC64x2048x2048t8_0p4_k200} at various times. The magnetic field undergoes pronounced inverse cascading. However, unlike the reference run R1, run R6 exhibits energy transfer to low spherical degree ($\ell \sim 1$), with $\EM(k) \approx 2 \times 10^{-9}$ compared to $\EM(k) \approx 6 \times 10^{-10}$ in run R1 at 
$\Tilde{t}=2\times 10^{-5}$. This energy transfer is less efficient than in runs R3 and R4. On the other hand, the inverse cascade remains limited to $\ell_0 \approx 50$, the same as in run R1 at $\Tilde{t}=2\times 10^{-5}$. Consequently, adopting periodic boundary conditions does not explain the formation of the large-scale magnetic field in magnetars.

In contrast to the reference run~R1, where the magnetic spectrum undergoes temporal decay,
run~R6 exhibits a stable value of $\EM^{\max}$ throughout its evolution.
The energy spectrum of run R6 aligns with the characteristics of fully helical MHD spectra; see the end of \Sec{subsec: cart coord} for details on a fully helical magnetic field and Figure 5 in \citet{brandenburg2020}.
Nonperiodic boundary conditions—such as potential, vertical, and perfect conductor boundaries—resulted in a partially helical field, explaining the decay in $\EM^{\max}$.
In contrast, the periodic boundary conditions in run~R6 facilitate the development of a fully helical magnetic field.

\section{Discussion}
\label{sec: discussion}

Studying the inverse cascade in NS crusts presents a significant challenge, as it remains a relatively unexplored topic in the existing literature.
A notable exception is the work of \citet{brandenburg2020}, which employed local simulations of the Hall cascade and identified the occurrence of an inverse cascade. However, these simulations, restricted to a cubic domain, are not adequate for capturing the complex and realistic properties of the NS crust. 

In contrast, \citet{gourgouliatos2020} explored the possibility of an inverse cascade in the NS crust but found no evidence of its occurrence. While an increase in the dipole component of the magnetic field was observed, the hallmark feature of an inverse cascade—a shift in the energy spectrum peak from smaller to larger scales—was absent. In their setup, magnetic energy was initially concentrated in multipoles within the range $\ell = 10 \ldots 20$, with minimal or zero contributions from other multipoles. Mode coupling, driven by the nonuniformity of coefficients and Hall nonlinearity, redistributed energy to neighboring multipoles and resulted in a shallow energy spectrum dominated by the initial scales. Instead of a spectral peak shifting to larger scales, energy redistributed to multipoles that initially carried negligible magnetic energy, with no discernible evidence of an inverse cascade manifesting. Over time, smaller-scale structures dissipated as Hall drift became less efficient, shaping their final energy spectra. The lack of magnetic helicity measurements in their study left a critical gap in understanding—specifically, whether the absence of an inverse cascade was due to the lack of magnetic helicity or the initial dominance of structures at scales comparable to the NS crust.

Despite these efforts, several critical factors remained unaddressed within a single comprehensive setup. These include high resolution simulations capable of resolving small-scale structures while incorporating the correct aspect ratio, spherical geometry, an initial causal spectrum (\Sec{sec: initial conditions}) designed to better capture conditions conducive to an inverse cascade, appropriate boundary conditions, and the consideration of time-dependent magnetic diffusivity—an essential factor influenced by the cooling processes of NSs.

To address these gaps, our research undertook a comprehensive investigation to evaluate the efficacy of the inverse cascade in the NS crust. Our work also aimed to understand its potential role in explaining the origin of the large-scale dipolar field observed in magnetars—an intriguing question with significant implications. A crucial requirement for a strong inverse cascade to occur is the presence of an initial helical magnetic field \citep{brandenburg2020}.
Nonhelical magnetic fields can also induce an inverse cascade via the Hall effect \citep{brandenburg2023Hall}, but it is weaker.
To produce an initially helical magnetic field, we developed a formalism for both Cartesian and spherical coordinates, allowing for systematic study using the \textsc{Pencil Code} and \texttt{MATINS}.

The slenderness of the NS crust adds complexity,
necessitating initial peak wavenumbers from small-scale structures on the order of $k_0 R \approx 200$ for the inverse cascade to occur.
Resolving these scales requires high-resolution simulations, which is challenging.
To overcome this, \texttt{MATINS} introduces additional wavenumbers in the radial direction, whereas the \textsc{Pencil Code} does not encounter this issue. \texttt{MATINS} lacks parallelization, making it less computationally powerful than the \textsc{Pencil Code}.

In our reference run R1, we constructed a model in Cartesian coordinates with an initial helical field dominated by small-scale structures, $k_0 R \approx 200$, while maintaining the correct aspect ratio of the NS crust (see \Sec{sec: ref run}). This setup was designed to optimally observe the inverse cascade, as the peak wavenumber shifted towards larger-scale structures within the crust's interior. However, the spectra revealed limitations: dissipation of the peak wavenumber (attributable to the use of nonperiodic boundary conditions that enforced a partially helical field), restriction of the cascade to $k_0 R \approx 30$, and negligible energy transfer to the largest-scale structure ($k R \approx 1$). To understand these phenomena, we explored the influence of geometrical coordinates, initial peak position, aspect ratio, and boundary conditions.

To explore the effects of geometry, we analyzed different coordinate systems in \Sec{sec: different coordinates}. Spherical coordinates were employed in the \textsc{Pencil Code} (run R2), with axis adjustments to avoid singularities, while cubed-sphere coordinates were used in \texttt{MATINS} (run R$\iota$5), which inherently bypass this issue. Despite these differences, the results were consistent, demonstrating that the choice between Cartesian and spherical domains has no significant effect when initial conditions are nearly identical. This observation aligns with findings by \citet{mitra2009}. Additionally, in \Sec{sec: effect of peak position}, we confirmed that small-scale initial spherical degrees ($\ell_0 \approx 200$) are essential for the occurrence of the inverse cascade.

Expanding our analysis, we examined the role of aspect ratio by exploring various crustal domain configurations (\Sec{sec: aspect ratio}). In the first case (Run R3), the crustal thickness was doubled. In the second case (Run R4), the crustal thickness was also doubled, but the domain in the angular directions $\theta/\pi$ and $\phi/\pi$ was halved, resulting in a nearly cubic geometry. Compared to the reference run R1,
three main differences were observed: First, the peak wavenumber $k_0 R$ continued shifting toward larger-scale structures and was no longer constrained to 30. Second, a significant transfer of energy toward the largest-scale structure ($k R = 1$) occurred.
Third, the efficiency of achieving a fully helical field was more pronounced, especially in run R4, highlighting the impact of geometrical configuration on the development of a fully helical field, even under nonperiodic boundary conditions.
These findings suggest that the thin-layer geometry of the NS crust hinders the inverse cascade, thereby limiting the formation of a large-scale magnetic field.

Building on these findings, we then investigated how different boundary conditions affect the spectral characteristics of the inverse cascade in \Sec{sec: Periodic BC}.
Our results indicated that, using periodic boundary conditions (run R6) rather than perfect conductor boundary conditions at the crust-core interface
or potential and vertical field boundary conditions at the surface, resulted in $\EM^{\max}$ being approximately constant. 
Periodic boundary conditions produce a fully helical magnetic field \citep[see][]{brandenburg2020},
unlike other boundary conditions, which result in a partially helical field \citep{brandenburg2017a}.
For a fully helical field, the constancy of the mean magnetic helicity density ensures that the spectral peak height $\EM^{\max}$ remains unchanged as the typical length scale $\xi$ increases and the mean magnetic energy density decreases (refer to the end of \Sec{subsec: cart coord}, for more details).

While periodic boundary conditions offer insights, they are not a realistic choice for studying the NS scenario.
Similarly, using perfect conductor boundary conditions, which expel the field from the NS core, and vertical field or potential boundary conditions, which prevent current flow outside the star, do not fully capture the realistic conditions of magnetic field evolution in a NS crust. Carefully addressing boundary conditions is crucial for gaining a better understanding of the inverse cascade in NS crusts. Although this level of detail is beyond the scope of this paper and not feasible with the numerical tools currently available, it remains an important consideration for future studies.

In this study, we disregarded the exact temperature-dependent microphysics and the stratification within the interior of a NS crust, anticipating that these factors would not significantly impact our findings regarding the inverse cascade.
The temperature-dependent microphysics is expected to reduce the magnetic diffusion parameter as the NS cools over time,
leading to a higher Lundquist number \citep[see Section~5.4 of][]{dehman2023a}. 
Our simulations have partially reflected this through a time-dependent magnetic diffusivity, as shown in \Eq{eq: eta pencil}.
Additionally, there is a radial dependence of $\Lu\propto(n_e\eta)^{-1}$; see \Eq{eq: Lu}.
\citet{gourgouliatos2020} considered the case of moderate stratification with $n_e\propto[1-(r-R)/H_e]^4$ and $H_e/R=0.0463$,
and used $\eta\propto n_e^{-2/3}$, so $\Lu\propto n_e^{-1/3}$ increased by a factor of about five from $r/R=1$ down to 0.9.
Using their approximation, \citet{brandenburg2020} found that $B_\text{rms}$ decays more slowly with depth, and $\xi$ increased only $\propto t^{1/3}$.
Nevertheless, the relation between magnetic dissipation and magnetic field strength was found to be remarkably similar to the unstratified case. 
Furthermore, \citet{cumming2004} demonstrated that at all depths, the transition from ohmic decay–dominated behavior to Hall-dominated behavior occurs at roughly the same time.

To ensure a Hall-dominant simulation and the occurrence of an inverse cascade, the Lundquist number must be significantly greater than one ($\mathrm{Lu} \gg 1$). We initialized our study with a Lundquist number on the order of a few hundred, simulating a magnetar-like scenario.
Consequently, our Hall-dominant simulations are representative of magnetic field evolution in middle-aged magnetars.
Since the aspect ratio limits the inverse cascade to field structures on the order of the crustal size, $k_0 R \approx 30$, we expect that considering a density (or radius) dependent Lundquist number will not change this conclusion.

In light of these insights, it is essential to consider realistic NS parameters, particularly the crustal thickness and its correct aspect ratio, when investigating the inverse cascade within a NS crust. While the inverse cascade cannot account for the formation of the large-scale dipolar field in magnetars—where the dipolar component remains on the order of $10^{12} \ldots 10^{13}$\,G—it does occur within the crust's interior at smaller scales.
These findings have significant implications for the evolution of NS magnetic fields. Our study reveals a field configuration characterized by a weak, large-scale dipolar component ($\approx 10^{12}$...$10^{13}\G$), along with dominant small-scale magnetic fields ($\approx 10^{14}$...$10^{15}\G$). The inverse cascade mechanism plays a crucial role by transferring energy from turbulent small scales to larger scales, effectively reducing small-scale dissipation. This energy transfer shapes the field's configuration and stability, influencing the crustal lattice dynamics and potentially impacting NS asteroseismology \citep{steiner2009,sotani2012,neill2023}, crustal failures \citep{perna2011,pons2011,dehman2020}, plastic flow within the star's interior \citep{lander15,lander19,gourgouliatos2021}, and may explain key observational properties of low-field magnetars and CCOs, which exhibit these characteristic magnetic features.

In summary, incorporating a helical magnetic field into the evolution of a NS's crust favors the occurrence of an inverse cascade, which reshapes the magnetic field over time.
This process influences key mechanisms that determine the star's physical properties and observational characteristics.

 \vspace{2mm}\noindent
{\em Software and data availability.}
The source code used for the simulations of this study,
the {\sc Pencil Code} \citep{pencil2021}, is freely available on
\url{https://github.com/pencil-code/}.
The simulation setups and corresponding input
and reduced out data are freely available on
\url{https://doi.org/10.5281/zenodo.14513354}; see also
\url{http://norlx65.nordita.org/~brandenb/projects/Reality-InvCasc-NS}
for easier access of the same data.

\begin{acknowledgements}
We thank the anonymous referee for their useful suggestions. We acknowledge the inspiring discussions with participants in the programs 
``Turbulence in Astrophysical Environments'' at the Kavli Institute for Theoretical Physics, Santa Barbara (NSF PHY-2309135),
and the Munich Institute for Astro-, Particle and BioPhysics (MIAPbP), which
is funded by the Deutsche Forschungsgemeinschaft (DFG, German Research Foundation) under Germany's Excellence Strategy -- EXC-2094 -- 390783311,
as well as with Jose A. Pons. We also acknowledge the visiting PhD fellow program at Nordita, which partially funded this research.  
This research was supported in part by the
Swedish Research Council (Vetenskapsr{\aa}det) under Grant No.\ 2019-04234,
the National Science Foundation under Grant No.\ AST-2307698 and the NASA ATP Award 80NSSC22K0825.
We also acknowledge funding from the Conselleria d'Educaci\'o, Cultura, Universitats i Ocupaci\'o de the Generalitat Valenciana
through grants CIPROM/2022/13 and ASFAE/2022/026 (with funding from NextGenerationEU PRTR-C17.I1), as well as the AEI grant PID2021-127495NB-I00 funded by MCIN/AEI/10.13039/501100011033. 
We acknowledge the allocation of computing resources provided by the
Swedish National Allocations Committee at the Center for
Parallel Computers at the Royal Institute of Technology in Stockholm.
\end{acknowledgements}

\bibliographystyle{aa} 
\bibliography{main} 

\appendix

\section{Magnetic formalism in 3D spherical coordinates}
\label{app: magnetic formalism sph}

In spherical coordinates, various formalism can be used to describe the magnetic field. In this context, we present the most common notations found in the literature. For any three-dimensional solenoidal vector field $\BB$, representing the magnetic field, we can always introduce the vector potential $\AAA$ such that
\begin{equation}
    \BB = \bnabla \times \AAA. 
    \label{eq: app B=nabla(A)}
\end{equation}
The magnetic field $\BB$ can be expressed using the poloidal $\Phi(\xx)$ and the toroidal $\Psi(\xx)$ scalar functions, based on the formalism of \citet{chandrasekhar1981}:
  \begin{eqnarray}
      \BB_\text{pol} &=&   \bnabla \times \AAA_\text{tor} = 
    \bnabla \times \big( \bnabla \times \Phi \boldsymbol{r}   \big),
      \nonumber\\
        \BB_\text{tor} &=&  \bnabla \times \AAA_\text{pol} =
        \bnabla\times \Psi \boldsymbol{r}.
        \label{eq: app poloidal toroidal field components}
  \end{eqnarray}
Using the notation of \citet{KR80} and \citet{geppert1991}, the two scalar functions can be expanded in a series of spherical harmonics:
\begin{eqnarray}
    \Phi(t,r,\theta,\phi) &=& \frac{1}{r}\sum_{\ell m} \Phi_{\ell m}(r,t) Y_{\ell m}(\theta,\phi),  \nonumber\\
    \Psi(t,r,\theta,\phi)  &=& \frac{1}{r}\sum_{\ell m} \Psi_{\ell m}(r,t) Y_{\ell m}(\theta,\phi), 
       \label{eq: app phi and Psi scalar functions}
  \end{eqnarray}
where $\ell= 1, ...,\ell_{max}$ is the degree and $m=-\ell,...,\ell$ the order of the multipole. 

The three components of the vector potential $\AAA$ in spherical coordinates are obtained by combining the poloidal and toroidal components:
\begin{eqnarray}
    A^r &=& \sum_{\ell m} \Psi_{\ell m}(r) Y_{\ell m} (\theta,\phi)~,  \nonumber\\
    A^{\theta} &=& \frac{1}{r ~ \sin\theta }  \sum_{\ell m} \Phi_{\ell m}(r) \frac{\partial Y_{\ell m} (\theta,\phi) }{\partial \phi}~,  \nonumber\\
     A^{\phi} &=&-  \frac{1}{r }  \sum_{\ell m} \Phi_{\ell m}(r) \frac{\partial Y_{\ell m} (\theta,\phi) }{\partial \theta}~. 
     \label{eq: vector potential components}
\end{eqnarray}
The three components of the magnetic field $\BB$ in the Newtonian limit are given by:  
\begin{eqnarray}
 B^{r} &=& \frac{1}{r^2} \sum_{\ell m} \ell(\ell+1)  \Phi_{\ell m}(r) Y_{\ell m}(\theta,\phi)~,   \nonumber\\
 B^{\theta} &=&   \frac{1}{r} \sum_{\ell m} \Phi_{\ell m}^\prime(r) \frac{\partial Y_{\ell m}(\theta,\phi)}{\partial \theta} \nonumber\\
 && +  \frac{1 }{r \sin\theta} \sum_{\ell m} \Psi_{\ell m}(r) \frac{\partial Y_{\ell m}(\theta,\phi)}{ \partial \phi} ~,
 \nonumber\\
  B^{\phi} &=&    - \frac{1}{r}\sum_{\ell m} \Psi_{\ell m}(r) \frac{\partial Y_{\ell m}(\theta,\phi)}{ \partial \theta} \nonumber\\
&&  + \frac{1}{r \sin\theta}  \sum_{\ell m}  \Phi_{\ell m}^\prime(r) \frac{\partial Y_{\ell m}(\theta,\phi)}{\partial \phi} ~. 
 \label{eq: magnetic field components spectral}
\end{eqnarray} 
Here, $\Phi^\prime_{\ell m} = \partial \Phi_{\ell m}/\partial r$, ignoring relativistic corrections. For the complete form, including relativistic corrections, refer to \citet{dehman2023a}. Using \EEq{eq: vector potential components} and \EEq{eq: magnetic field components spectral}, one can derive the expressions for the spectral magnetic helicity \Eq{eq: spectral magnetic helicity}, the spectral magnetic energy \Eq{eq: energy spectrum MATINS}, and the spectral realizability condition \Eq{eq: realisability l}. 

\end{document}